\documentclass[sigconf,screen,authorversion]{acmart}

\usepackage[utf8]{inputenc}
\usepackage[T1]{fontenc}
\usepackage{microtype}

\usepackage{braket}
\usepackage{subfigure}
\usepackage{tabularx}
\usepackage{multirow}
\usepackage{pifont}
\usepackage{listings}

\setcopyright{acmcopyright}
\acmPrice{}
\acmDOI{10.1145/3445814.3446750}
\acmYear{2021}
\copyrightyear{2021}
\acmSubmissionID{asplos21main-p936-p}
\acmISBN{978-1-4503-8317-2/21/04}
\acmConference[ASPLOS '21]{Proceedings of the 26th ACM International Conference on Architectural Support for Programming Languages and Operating Systems}{April 19--23, 2021}{Virtual, USA}
\acmBooktitle{Proceedings of the 26th ACM International Conference on Architectural Support for Programming Languages and Operating Systems (ASPLOS '21), April 19--23, 2021, Virtual, USA}

\begin{document}

\title[Logical Abstractions for Noisy Variational Quantum Algorithm Simulation]{Logical Abstractions for Noisy Variational \\ Quantum Algorithm Simulation}

\author{Yipeng Huang}
\email{yipeng.huang@rutgers.edu}
\orcid{0000-0003-3171-6901}
\affiliation{%
  \institution{Rutgers University}
  \streetaddress{110 Frelinghuysen Road}
  \city{New Brunswick}
  \state{NJ}
  \country{United States}
  \postcode{08854-8019}
}

\author{Steven Holtzen}
\email{sholtzen@cs.ucla.edu}
\affiliation{%
  \institution{University of California, Los Angeles}
  \streetaddress{404 Westwood Plaza, Engineering VI}
  \city{Los Angeles}
  \state{CA}
  \country{United States}
  \postcode{90095-1596}
}

\author{Todd Millstein}
\email{todd@cs.ucla.edu}
\affiliation{%
  \institution{University of California, Los Angeles}
  \streetaddress{404 Westwood Plaza, Engineering VI}
  \city{Los Angeles}
  \state{CA}
  \country{United States}
  \postcode{90095-1596}
}

\author{Guy Van den Broeck}
\email{guyvdb@cs.ucla.edu}
\affiliation{%
  \institution{University of California, Los Angeles}
  \streetaddress{404 Westwood Plaza, Engineering VI}
  \city{Los Angeles}
  \state{CA}
  \country{United States}
  \postcode{90095-1596}
}

\author{Margaret Martonosi}
\email{mrm@princeton.edu}
\affiliation{%
 \institution{Princeton University}
  \streetaddress{35 Olden Street}
  \city{Princeton}
  \state{NJ}
  \country{United States}
  \postcode{08540-5233}
}


\begin{abstract}






Due to the unreliability and limited capacity of existing quantum computer prototypes,
quantum circuit simulation continues to be a vital tool for validating next generation quantum computers and for studying variational quantum algorithms,
which are among the leading candidates for useful quantum computation.
Existing quantum circuit simulators do not address the common traits of variational algorithms, namely:
1) their ability to work with noisy qubits and operations,
2) their repeated execution of the same circuits but with different parameters, and
3) the fact that they sample from circuit final wavefunctions to drive a classical optimization routine.
We present a quantum circuit simulation toolchain based on logical abstractions targeted for simulating variational algorithms.
Our proposed toolchain encodes quantum amplitudes and noise probabilities in a probabilistic graphical model,
and it compiles the circuits to logical formulas that support efficient repeated simulation of and sampling from quantum circuits for different parameters.
Compared to state-of-the-art state vector and density matrix quantum circuit simulators,
our simulation approach offers greater performance when sampling from noisy circuits with at least eight to 20 qubits and with around 12 operations on each qubit,
making the approach ideal for simulating near-term variational quantum algorithms.
And for simulating noise-free shallow quantum circuits with 32 qubits, our simulation approach offers a $66\times$ reduction in sampling cost versus quantum circuit simulation techniques based on tensor network contraction.
\end{abstract}

\ccsdesc[500]{Computer systems organization~Quantum computing}
\ccsdesc[500]{Mathematics of computing~Probabilistic inference problems}
\ccsdesc[300]{Computing methodologies~Knowledge representation and reasoning}
\ccsdesc[100]{Mathematics of computing~Bayesian networks}
\ccsdesc[100]{Mathematics of computing~Decision diagrams}

\keywords{quantum circuits, Bayesian networks, conjunctive normal form, knowledge compilation, exact inference, simulation}


\maketitle


\vspace{-20pt}
\begin{figure}[h]
\begin{minipage}[t]{0.33\linewidth}
\centering
\caption*{Before}
\includegraphics[width=\linewidth]{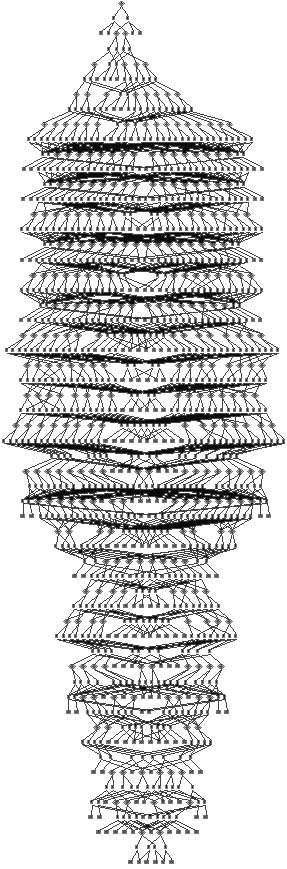}
\end{minipage}
\hfill
\begin{minipage}[t]{0.66\linewidth}
\centering
\caption*{After}
\includegraphics[width=.66\linewidth]{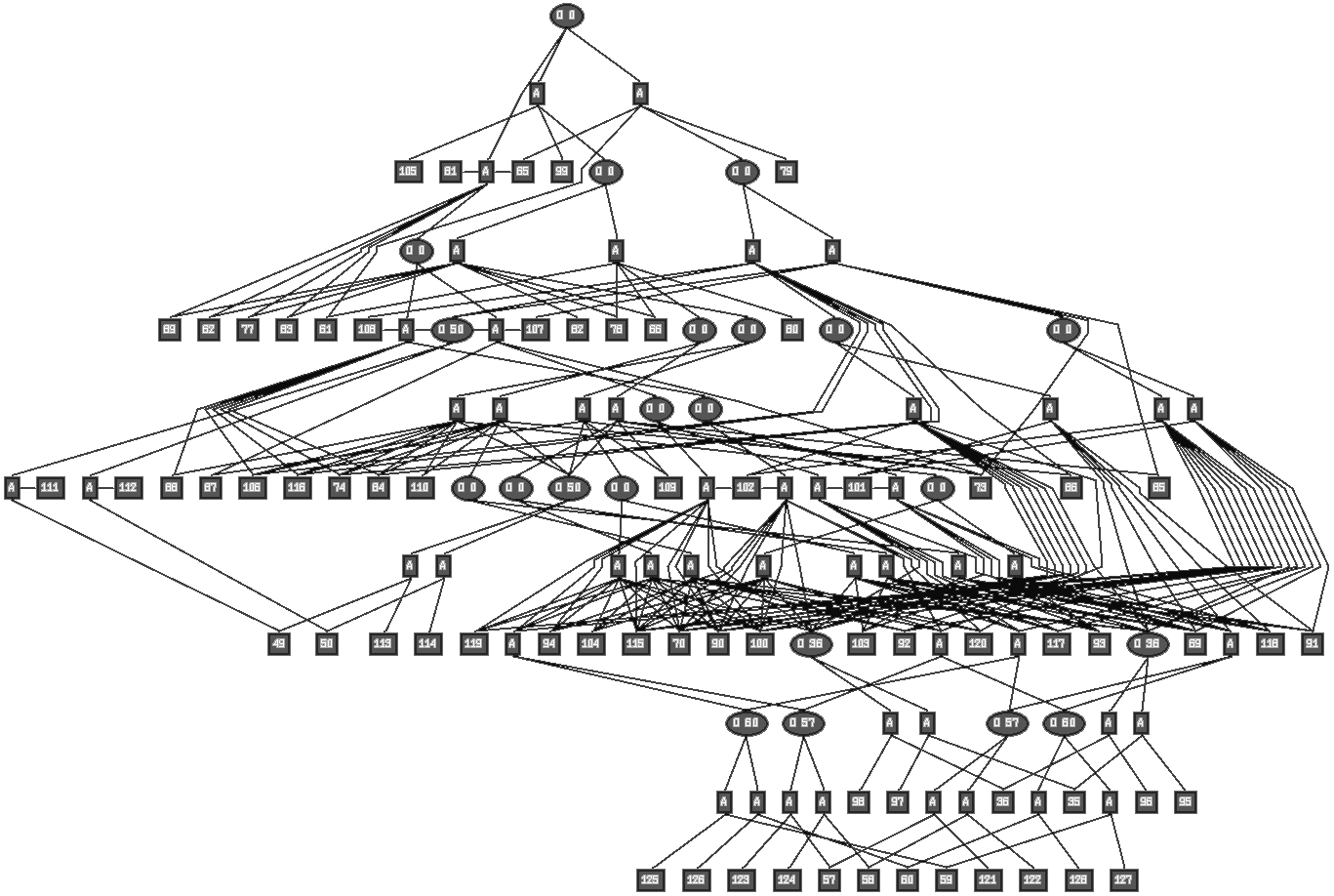}
\vspace{-5pt}
\caption{
Equivalent knowledge compilation representations of a 4-qubit noisy QAOA quantum circuit.
In this work we calculate and sample amplitudes from arithmetic circuits (ACs) representing noisy quantum circuits.
To the left, direct compilation results in ACs where qubit states ordered in time progresses from top to bottom.
Above, optimizations such as logical minimization, qubit state reordering, and eliding internal qubit states reduce the size of the AC.
The reduced but equivalent representation leads to more efficient simulation and sampling.
}
\label{fig:vcg}
\end{minipage}
\end{figure}

\section{Introduction}

Consensus among quantum systems researchers is that variational algorithms are among the most important near-term applications of quantum computing~\cite{Preskill2018quantumcomputingin,NAP25196}.
Variational algorithms work by using a classical computer to train for optimal parameters that minimize a function evaluated by a quantum computer.
Examples include the quantum approximate optimization algorithm (QAOA)~\cite{farhi_qaoa,farhi2016quantum} and the variational quantum eigensolver (VQE)~\cite{vqe} algorithm for physical simulations.
Unlike prominent quantum algorithms such as Shor's factoring~\cite{doi:10.1137/S0097539795293172} and Grover's search~\cite{Grover:1996:FQM:237814.237866} algorithms, variational algorithms can extract useful computation out of noisy intermediate-scale quantum (NISQ) computer prototypes, which only support unreliable operations on a limited number of qubits (the fundamental unit of quantum computing).

A further consequence of the limited capacity, reliability, and endurance of existing quantum prototypes is that simulation using classical computers continues to be a critical research tool~\cite[Chapter 6.3]{NAP25196}.
Classical computer simulations of quantum algorithm circuits are important for developing new quantum algorithms and for validating results from quantum prototypes.

Thus far, the most advanced quantum circuit simulators are not geared for simulating important variational quantum algorithms.
Quantum computing research would benefit from a simulator that supports variational algorithms specifically, which would require a simulator that
1) supports simulating the effect of noise,
2) efficiently supports repeated simulations with different parameters,
3) offers an ability to sample from the output of the quantum computer, and
4) excels at simulating quantum circuits with many qubits and relatively few operations per qubit.
Unfortunately, leading quantum circuit simulators have instead focused on simulations that establish the point at which limitations of classical computing give way to quantum computers having an advantage, a milestone termed quantum supremacy~\cite{google_quantum_supremacy,Harrow,markov2018quantum,Terhal}.
Such simulators can only simulate quantum circuits that are ideal (noise-free), and they cannot reuse computation results across simulation runs.
The clear mismatch between the requirements for variational algorithm versus supremacy simulations have resulted in simulators that do not adequately support important variational workloads.

The key insight of our paper is that \emph{knowledge compilation}---a technique for efficient repeated inference originating in artificial intelligence research~\cite{darwiche2009modeling,darwiche2002knowledge}---can serve as the basis for a quantum circuit simulation toolchain geared for variational algorithms.
In a knowledge compilation approach to performing inference, knowledge about probabilistic relationships between events is first encoded in a graphical model such as a Bayesian network~\cite{darwiche2009modeling,Koller:2009:PGM:1795555,pearl2014probabilistic}.
The knowledge compilation techniques convert Bayesian networks into minimized representations of logical formulas called arithmetic circuits (ACs, Figure~\ref{fig:vcg}) that enable repeated inference and sampling queries with different parameters and new choices for inference outcomes~\cite{Chavira:2008:PIW:1342435.1342783}.
These features of the knowledge compilation approach---namely, 1) the ability to represent and manipulate probabilistic information, 2) the ability to compile probabilistic model structural information into minimized formats, 3) the ability to efficiently sample from the same model but for varying parameters and evidence---match well with the requirements for variational quantum algorithm simulation.

We built a toolchain to test this idea of using knowledge compilation for variational algorithms quantum circuit simulation. Our toolchain consists of:
\begin{enumerate}

\item A front-end for converting noisy quantum circuits (specified in Google's Cirq framework\footnote{\url{https://github.com/quantumlib/Cirq}}) to complex-valued Bayesian networks~\cite{2017arXiv170800006B,boixo2017simulation,doi:10.1137/050644756,2015NJPh...17g3020P,tucci1995quantum,5708827}, which we extend to correctly encode quantum noise mixtures and channels.
Compared to conventional quantum circuits where complex-valued quantum amplitudes and real-valued noise probabilities are treated separately,
the Bayesian network encoding unifies quantum states and noise events in a single representation.

\item A compiler that converts Bayesian networks representing noisy quantum circuits into conjunctive normal form (CNF) logic formulas.
The CNFs encode the quantum circuits' structural information:
the sets of logic variable assignments that satisfy the CNF correspond to all sets of qubit state assignments that are consistent with the original quantum circuit's semantics.
This structural information can be reused across simulations independently of quantum amplitude and noise probability parameters, which vary across simulations, which is a key benefit over prior simulation techniques.

\item A compiler that converts CNFs to ACs. 
An AC enumerates and assigns a weight value to each set of variable assignments that satisfy a logical formula~\cite{darwiche2009modeling}.
Summing the weights across all qubit state assignments results in the output amplitudes that we seek to find in the quantum circuit simulation task.
The compiler can factor away the variables that represent intermediate qubit states, thereby enabling the quantum circuit simulator to find the probability amplitude for an outcome without incurring the cost of finding the amplitudes of intermediate qubit states.
The ACs also enable a Markov chain Monte Carlo procedure for sampling sets of qubit outcomes according to their measurement probability.


\end{enumerate}

We validate our compilation and simulation approach for both noise-free and noisy quantum circuits,
demonstrating correct results for a suite of quantum algorithms including Deutsch-Jozsa, Bernstein-Vazirani, hidden shift, quantum Fourier transform, Shor's, and Grover's algorithms.

We benchmark the performance of our simulator for sampling outputs for a QAOA algorithm for Max-Cut and a VQE algorithm for finding the minimum energy configuration of a 2D Ising model.
Compared to state-of-the-art simulators for both ideal and noisy quantum circuits, our simulator excels at sampling from circuits with at least eight to 20 qubits and with around 12 operations per qubit---a range of qubit counts and operations that includes many meaningful variational algorithm problems.
And for simulating ideal shallow quantum circuits with 32 qubits, our simulation approach offers a $66\times$ reduction in sampling cost versus simulators based on tensor network contraction.
The advantages are due to the more compact representation, the circuit minimization and memoization capabilities of our approach, and due to the storage costs for conventional simulators based on matrix representations.
The improved simulation performance facilitates studying variational algorithms and validating prototype quantum computer results in the NISQ era of quantum computing.

\begin{figure*}[t]
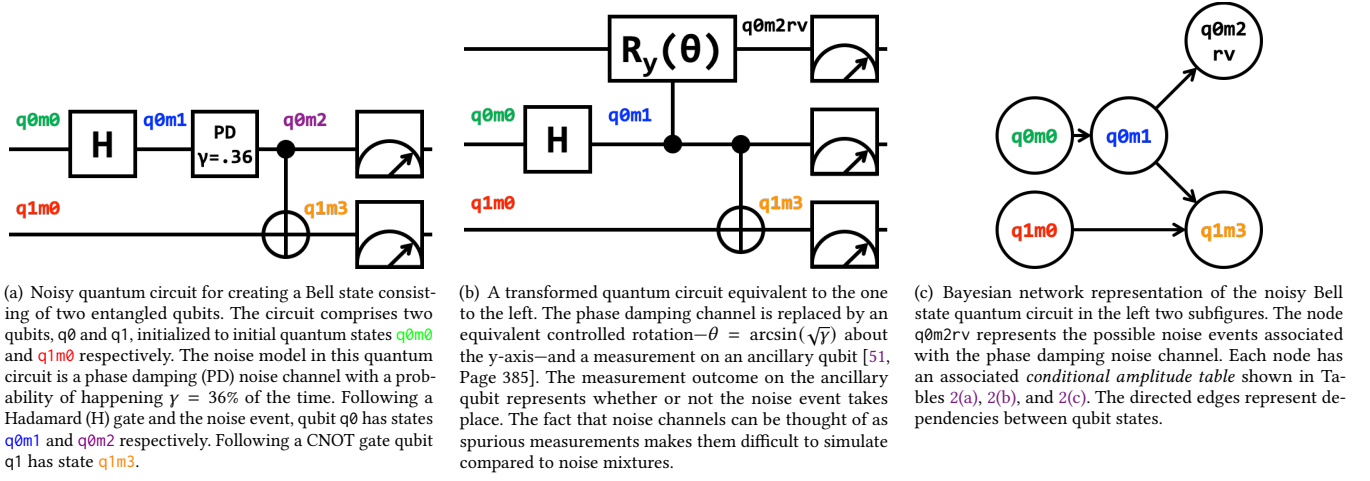

     \centering
\subfigure[
Noisy quantum circuit for creating a Bell state consisting of two entangled qubits.
The circuit comprises two qubits, \texttt{q0} and \texttt{q1}, initialized to initial quantum states \textcolor{green}{\texttt{q0m0}} and \textcolor{red}{\texttt{q1m0}} respectively.
The noise model in this quantum circuit is a phase damping (PD) noise channel with a probability of happening $\gamma=36\%$ of the time.
Following a Hadamard (H) gate and the noise event, qubit \texttt{q0} has states \textcolor{blue}{\texttt{q0m1}} and \textcolor{violet}{\texttt{q0m2}} respectively.
Following a CNOT gate qubit \texttt{q1} has state \textcolor{orange}{\texttt{q1m3}}.
]{\label{fig:noisy_bell}\includegraphics[width=0.32\textwidth]{figures/noisy_bell.pdf}}
     \hfill
\subfigure[
A transformed quantum circuit equivalent to the one to the left.
The phase damping channel is replaced by an equivalent controlled rotation---$\theta=\mathrm{arcsin}(\sqrt{\gamma})$ about the y-axis---and a measurement on an ancillary qubit~{\cite[Page~385]{Nielsen:2011:QCQ:1972505}}.
The measurement outcome on the ancillary qubit represents whether or not the noise event takes place.
The fact that noise channels can be thought of as spurious measurements makes them difficult to simulate compared to noise mixtures.
]{\label{fig:noisy_bell_transformed}\includegraphics[width=0.32\textwidth]{figures/noisy_bell_transformed.pdf}}
     \hfill
\subfigure[
Bayesian network representation of the noisy Bell state quantum circuit in the left two subfigures.
The node \texttt{q0m2rv} represents the possible noise events associated with the phase damping noise channel.
Each node has an associated \emph{conditional amplitude table} shown in Tables~\ref{tab:cat_hadamard},~\ref{tab:cat_phase_damping}, and~\ref{tab:cat_cnot}.
The directed edges represent dependencies between qubit states.
]{\makebox[0.32\textwidth]{\label{fig:noisy_bn}\includegraphics[width=.2\linewidth]{figures/noisy_bn.pdf}}}
        \caption{Noisy Bell state quantum circuit and its Bayesian network representation.}
        \label{fig:noisy_bell_figure}
\end{figure*}

\section{Background on Quantum Circuit Simulation and Variational Algorithms}

We summarize necessary background on the representation of noise-free and noisy quantum states, on simulating quantum circuits, and on variational quantum algorithms.
\subsection{Ideal Quantum Circuit Simulation}
\label{sec:ideal_simulation}

Quantum circuit simulation entails using a classical computer to calculate the outputs of a quantum computer.
Quantum circuit simulation is useful for discovering new quantum algorithms,
validating the execution of quantum programs on unreliable quantum hardware~\cite{google_quantum_supremacy,PhysRevA.100.032328,resch2019benchmarking,huang_et_al:OASIcs:2019:10196,huang_isca}, and understanding the limitations of classical and quantum computation~\cite{google_quantum_supremacy,Harrow,markov2018quantum,Terhal}.
To understand quantum circuit simulation, first we have to understand the representation of quantum states and operations.

\subsubsection{The State Vector Representation for Quantum Pure States}

The fundamental unit of computation in a quantum computer is a qubit.
A single qubit has a state represented by the vector $ \alpha\ket{0} + \beta\ket{1} = \begin{bmatrix}\alpha,\beta\end{bmatrix}^\intercal$,
where $\ket{0}$ and $\ket{1}$ are orthonormal standard basis vectors and $\alpha$ and $\beta$ are complex-valued amplitudes.
It is required that $|\alpha|^2 + |\beta|^2 = 1$.
The state of $N$ qubits is represented by a state vector that has size $2^N$.
For example, two qubits have a state represented by the vector $ \alpha\ket{00} + \beta\ket{01} + \gamma\ket{10} + \delta\ket{11} = \begin{bmatrix}\alpha,\beta,\gamma,\delta\end{bmatrix}^\intercal$,
where again $|\alpha|^2 + |\beta|^2 + |\gamma|^2 + |\delta|^2 = 1$.

Using Figure~\ref{fig:noisy_bell} as an example, if qubit state $\textcolor{green}{\texttt{q0m0}}=\ket{0}$ and $\textcolor{red}{\texttt{q1m0}}=\ket{0}$, then the state vector of the two qubits is found using the tensor product (denoted $\otimes$):
\begin{align*}
    \textcolor{green}{\texttt{q0m0}} \otimes \textcolor{red}{\texttt{q1m0}} =
    \ket{0} \otimes \ket{0} =
    \begin{bmatrix}1 \\ 0 \end{bmatrix} \otimes \begin{bmatrix}1 \\ 0\end{bmatrix}
    = \begin{bmatrix} 1 \\ 0 \\ 0 \\ 0 \end{bmatrix}
    = \ket{00}
\end{align*}

\subsubsection{The Unitary Matrix Representation of Quantum Gates and Circuits}

Quantum computation proceeds by applying quantum gates on quantum states encoded on qubits.
Quantum gates are represented by norm-preserving unitary matrices, that is, matrices that ensure that the sums of squares of amplitudes remain 1.
For example one important quantum gate is the Hadamard gate which has a unitary matrix representation of:
\begin{align*}
    H=
    \begin{bmatrix}
        \frac{1}{\sqrt{2}} & \frac{1}{\sqrt{2}} \\
        \frac{1}{\sqrt{2}} & -\frac{1}{\sqrt{2}}
    \end{bmatrix}
\end{align*}

When quantum gates act on pure states, the resulting state is found via matrix vector multiplication of unitary matrices and state vectors.
Using Figure~\ref{fig:noisy_bell} as an example, the qubit state \textcolor{blue}{\texttt{q0m1}} after the Hadamard gate is:

\begin{align}
    \textcolor{blue}{\texttt{q0m1}} = 
    H\textcolor{green}{\texttt{q0m0}} =
    \begin{bmatrix}
        \frac{1}{\sqrt{2}} & \frac{1}{\sqrt{2}} \\
        \frac{1}{\sqrt{2}} & -\frac{1}{\sqrt{2}}
    \end{bmatrix}
    \begin{bmatrix}1 \\ 0 \end{bmatrix} =
    \begin{bmatrix}\frac{1}{\sqrt{2}} \\ \frac{1}{\sqrt{2}} \end{bmatrix} = \frac{1}{\sqrt{2}}\ket{0} + \frac{1}{\sqrt{2}}\ket{1}
    \label{eq:q0m1}
\end{align}


\subsection{Noisy Quantum Circuit Simulation}
\label{sec:noisy_simulation}

In contrast to simulating noise-free ideal quantum circuits, simulating realistic quantum circuits requires the ability to represent noisy quantum mixed states and the ability to model various non-ideal effects.
Existing prototype quantum computers are unreliable due to various non-ideal effects. The effects include environmental disturbance of delicate quantum states leading to decoherence, imprecise application of operations, and measurement error.
The outcome of quantum states in a noisy quantum circuit varies depending on whether noise events takes place, so a greater amount of information is needed to account for all the possibilities, thereby making the task of simulating noisy circuits harder than simulating ideal ones.
To understand this challenge we introduce the representation of noisy quantum states and models of quantum noise.

\subsubsection{The Density Matrix Representation of Quantum Mixed States}

Density matrices represent noisy quantum states as probabilistic ensembles of pure states.
A density matrix $\rho$ for pure states $\ket{\psi}$ has the form:
$\rho = \sum_{j} p_j \ket{\psi_j}\bra{\psi_j}$,
where $p_j$ is the probability that the mixed state is the pure state $\ket{\psi_j}$,
$\bra{\psi_j}$ is the conjugate transpose of $\ket{\psi_j}$,
and $\sum_{j} p_j = 1$.
Using Figure~\ref{fig:noisy_bell} as an example, we may wish to have a density matrix representation of \textcolor{blue}{\texttt{q0m1}} in preparation for calculating the effect of quantum noise on that qubit state. The density matrix representation of \textcolor{blue}{\texttt{q0m1}} from Equation~\ref{eq:q0m1} is:

\begin{align}
\begin{split}
    \rho_{\textcolor{blue}{\texttt{q0m1}}}
&=
    \Big(\frac{1}{\sqrt{2}}\ket{0} + \frac{1}{\sqrt{2}}\ket{1}\Big) \Big(\frac{1}{\sqrt{2}}\bra{0} + \frac{1}{\sqrt{2}}\bra{1}\Big) \\
&=
\begin{bmatrix}\frac{1}{\sqrt{2}} \\ \frac{1}{\sqrt{2}} \end{bmatrix} \begin{bmatrix}\frac{1}{\sqrt{2}}& \frac{1}{\sqrt{2}} \end{bmatrix} = \begin{bmatrix}\frac{1}{2}&\frac{1}{2} \\ \frac{1}{2}&\frac{1}{2} \end{bmatrix}
\end{split}
\label{eq:density_matrix}
\end{align}

\subsubsection{The Kraus Operator Representation of Quantum Noise Channels}
\label{sec:noise_mixtures_channels}

\begin{table}
\caption{Summary of canonical quantum noise models.}
\label{tab:noise_models}
\small
\begin{tabularx}{\linewidth}{p{.07cm} p{1.4cm} p{2.8cm} p{2.8cm}}
\toprule
& & Quantum noise mixtures & Quantum noise channels \\\midrule
\multirow{6}{*}{\rotatebox[origin=c]{90}{Noise effects}}
& Pauli-X type & Bit flip noise & Amplitude damping\newline noise (related to T1 time) \\
& Pauli-Z type & Phase flip noise & Phase damping\newline noise (related to T2 time) \\
& Combinations & Symmetric / asymmetric depolarizing noise & Generalized amplitude damping \\\midrule
\multicolumn{2}{l}{Sim. technique} & Can model as probabilistic ensembles of state vectors & Requires density matrix representation \\\bottomrule
\end{tabularx}
\end{table}

Quantum noise can be modeled as a quantum noise channel $\mathcal{E}$, which acts on a quantum mixed state $\rho$ to create a new mixed state:
$\mathcal{E}(\rho) = \sum_k E_k \rho E_k^\intercal$,
where the Kraus operators $E$ represent different effects on the quantum state due to the noise channel.

Important noise channel types are listed in Table~\ref{tab:noise_models}~\cite[Chapter 8.3]{Nielsen:2011:QCQ:1972505}.
The table shows that noise models can be classified along several dimensions.
The first dimension is in the type of effect the noise has on the quantum state: Pauli-X type noises disturb the quantum basis state, while Pauli-Z type noises disturb the phase. There are also combinations of these types of noise. The second dimension of classification is in terms of whether density matrices are needed to model the noisy states.
In this work we consider all of these types of noise.

For example, one type of quantum noise channel is phase damping noise, which has the Kraus operators:
\begin{align*}
E_0 = \begin{bmatrix}
1&0\\
0&\sqrt{1-\gamma}
\end{bmatrix},
\qquad
E_1 = \begin{bmatrix}
0&0\\
0&\sqrt{\gamma}
\end{bmatrix}
\end{align*}
where $\gamma$ is a probability parameter describing the strength of the noise channel.
Using Figure~\ref{fig:noisy_bell} as an example, the density matrix representation of qubit state \textcolor{violet}{\texttt{q0m2}} after a phase damping channel (with $\gamma=.36$) acts on \textcolor{blue}{\texttt{q0m1}} from Equation~\ref{eq:density_matrix} is:

\begin{align*}
    \rho_{\textcolor{violet}{\texttt{q0m2}}} = \mathcal{E}(\rho_{\textcolor{blue}{\texttt{q0m1}}}) =
    \begin{bmatrix}\frac{1}{2}&\frac{0.8}{2}\\\frac{0.8}{2}&\frac{1}{2}\end{bmatrix}
\end{align*}

Finally, the two-qubit CNOT gate has a unitary matrix representation of:
\begin{align*}
CNOT =
\begin{bmatrix}
1 & 0 & 0 & 0 \\
0 & 1 & 0 & 0 \\
0 & 0 & 0 & 1 \\
0 & 0 & 1 & 0 \\
\end{bmatrix}
\end{align*}
such that the final density matrix is:
\begin{align}
\begin{split}
    \rho_{\textcolor{violet}{\texttt{q0m2}}}\otimes\rho_{\textcolor{orange}{\texttt{q1m3}}} &=
    CNOT(\rho_{\textcolor{violet}{\texttt{q0m2}}}\otimes\rho_{\textcolor{red}{\texttt{q1m0}}})CNOT^\intercal\\
    &=
\begin{bmatrix}
\frac{1}{2} & 0 & 0 & \frac{0.8}{2} \\
0 & 0 & 0 & 0 \\
0 & 0 & 0 & 0 \\
\frac{0.8}{2} & 0 & 0 & \frac{1}{2} \\
\end{bmatrix}
\end{split}
\label{eq:final}
\end{align}

Outside of these canonical quantum noise mixture and channel models, other types of operational error---such as imprecise gate operations and measurement error---can be modeled as one of the canonical noise models in conjunction with an otherwise ideal operation or measurement.

\begin{figure*}[t]{}
\centering
\includegraphics[width=\linewidth]{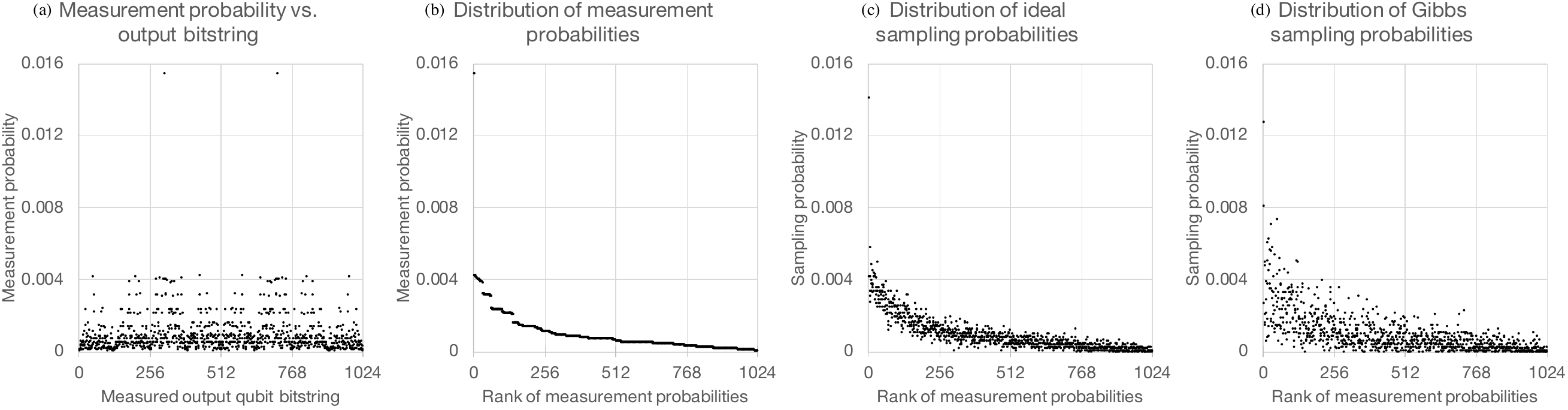}
\caption{The probability distribution for output qubit measurements in variational quantum algorithms is sharply peaked. A few qubit bitstrings dominate the output probability distribution for this 10-qubit quantum circuit performing QAOA for Max-Cut.
Since a few bitstrings dominate the outcomes, sampling the outcomes is more efficient than finding the full probability distribution.
}
\label{fig:noise_qaoa_1024}
\end{figure*}

\begin{figure*}[t]{}
\centering
\includegraphics[width=\linewidth]{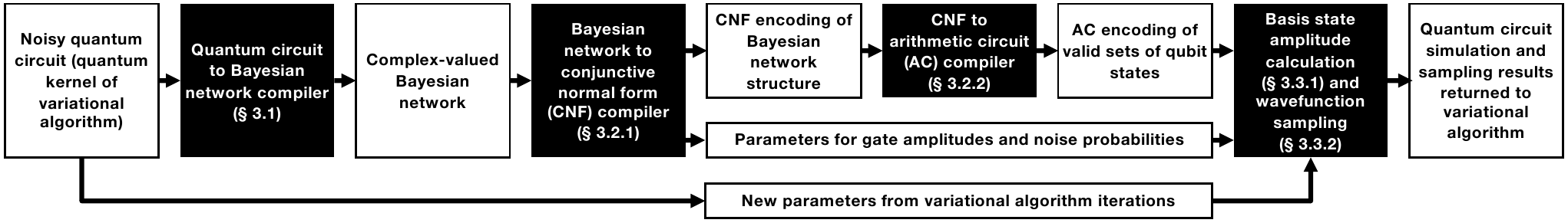}
\caption{Toolchain for noisy quantum algorithm simulation via knowledge compilation of probabilistic program representations.
}
\label{fig:toolchain}
\end{figure*}

\subsection{Near-Term Variational Quantum Algorithms as Target Simulation Workload}

We focus on quantum circuit simulation for algorithms that are candidates for near-term useful quantum computation.
Such algorithms are designed to run on near-term noisy, intermediate-scale quantum (NISQ) computers~\cite{10.1007/978-3-030-14082-3_7,bravoprieto2019variational,huang2019nearterm,PhysRevA.101.010301,NAP25196,Preskill2018quantumcomputingin}.
Specifically, we evaluate our simulation approach on two representative NISQ algorithms:
one is a quantum approximate optimization algorithm (QAOA) for Max-Cut~\cite{arute2020quantum,farhi_qaoa,farhi2016quantum,szegedy2019qaoa},
and the other is a variational quantum eigensolver (VQE) for an Ising model physics simulation~\cite{Barahona_1982,vqe}.

These hybrid quantum-classical algorithms rely on a classical computer running an optimization routine such as the Nelder-Mead method to find optimal parameters for a quantum circuit.
The quantum computer serves only to find an objective function from the system under study to guide the overall optimization loop.
The quantum circuit parameters that minimize the objective function encode the desired algorithm results.



The quantum circuits involved in these important variational algorithms have distinct traits, and so simulating these circuits is also a distinct challenge.
Compared to the quantum circuits involved in other algorithms, variational quantum algorithms: \begin{enumerate}
    \item do not rely on error-corrected ideal qubits and operations, and are therefore sensitive to the reliability and noise characteristics of the underlying hardware;
    \item require repeated execution or simulation of the same circuit but with different parameters;
    \item use circuits that are wide but shallow (\textit{i.e.}, they use many qubits but perform relatively few operations on those qubits);
    \item rely on the quantum computer or simulator to sample from the final quantum wavefunction, which have measurement probability distributions that are sharply peaked (Figure~\ref{fig:noise_qaoa_1024}a).
\end{enumerate}
These traits also set variational algorithm circuits apart from those in random circuit sampling circuits, which have thus far been the focus for quantum circuit simulators.
In order to accelerate the development of these NISQ variational algorithms, researchers need high performance and efficient simulators that specifically support NISQ algorithms.


\section{Noisy Quantum Algorithm Simulation via Knowledge Compilation}
\label{sec:toolchain}


Our approach to quantum program simulation involves a series of program transformations that enables more efficient simulation of variational quantum algorithms.
We adapt techniques from classical Bayesian inference by converting the simulation problem into one of repeated inference and sampling from a probabilistic graphical model (PGM).
The toolchain comprises three main stages, corresponding to the special traits of variational algorithms (Figure~\ref{fig:toolchain}).
\begin{enumerate}
    \item \textbf{Conversion of noisy quantum circuits to Bayesian networks}, (a kind of PGM).
    This program translation combines in a single representation the two types of values in noisy quantum circuits:
    the real-valued probabilities associated with noise events and the complex-valued quantum amplitudes associated with qubits and gates.
    The unified representation enables more direct manipulation and simulation of noise effects. (Section~\ref{sec:qc_to_bn})
    \item \textbf{Knowledge compilation of the Bayesian networks}.
    This step borrows techniques originating in artificial intelligence research meant for efficient repeated inference on PGMs.
    Our toolchain compiles the structure of Bayesian networks into conjunctive normal form (CNF) logic formulas to separate the quantum circuit structure from gate and noise parameters (Section~\ref{sec:bn_to_cnf}).
    The toolchain then compiles the CNFs to arithmetic circuits that can be reused across quantum program simulations with different parameters as needed in variational algorithms (Section~\ref{sec:cnf_to_ac}).
    Ours is the first work to demonstrate such reuse of computational results across simulation runs.
    \item \textbf{Gibbs sampling on the compiled PGM representation}.
    Following the previous transformations, the task of finding the amplitude associated with a given assignment of qubit values becomes equivalent to the task of finding the probability of a given set of evidence in a Bayesian network.
    For the wide but shallow circuits typically found in variational algorithms, doing this type of simulation is more efficient than finding full state vectors (Section~\ref{sec:inference}).
    Compiling to arithmetic circuits further enables the simulator to use a Markov chain Monte Carlo method to draw measurement outcomes in the same way a prototype quantum computer would (Section~\ref{sec:sampling}).

\end{enumerate}
We discuss these program transformations using a detailed example in the following subsections.


\subsection{Converting Noisy Quantum Circuits to Bayesian Networks}
\label{sec:qc_to_bn}


The first stage of our program transformation is to convert noisy quantum circuits into complex-valued Bayesian networks.
We perform this transformation to combine the real-valued probabilities associated with quantum noise events with the complex-valued quantum amplitudes associated with qubit states and gates.
Such a transformation is possible because quantum programs and Bayesian networks are both inherently probabilistic;
with suitable changes to the latter's semantics they can represent ideal quantum circuits with no loss of generality~\cite{2017PhRvX...7c1021A,boixo2017simulation,Henson_2014,tucci1995quantum,5708827}.

In classical inference, Bayesian networks are a basic type of PGM~\cite{darwiche2009modeling,Koller:2009:PGM:1795555,Korb:2010:BAI:1941985,pearl2014probabilistic}.
They consist of network nodes that in the classical setting represent probabilistic random variables.
Directed edges in Bayesian networks represent conditional dependence.
Additionally, each node is associated with a conditional probability table that describes the conditional probability of that node's variable given knowledge about that variable's dependencies.

In the quantum setting, Bayesian network nodes represent qubit states, and directed edges represent how qubit states depend on preceding qubit states.
In contrast to the quantum circuits representation that dominates quantum computing research and teaching~\cite{Deutsch_Networks,Kaye:2007:IQC:1206629,mermin2007quantum,Nielsen:2011:QCQ:1972505},
a Bayesian network representation of a quantum program emphasizes the graphical structure of dependencies between the qubit states and operations on qubits.

In this work, we extend these quantum PGMs~\cite{2017PhRvX...7c1021A,boixo2017simulation,Henson_2014,tucci1995quantum,5708827} to represent probabilistic events associated with noisy quantum operations.
Figure~\ref{fig:noisy_bell_figure} shows the transformation of a noisy quantum circuit for creating Bell states to its corresponding Bayesian network representation.
We'll be using this minimal example throughout Section~\ref{sec:toolchain}.

\subsubsection{Encoding Ideal Qubits \& Operations}

Quantum Bayesian networks encode the unitary matrices associated with quantum gates as \emph{conditional amplitude tables}, which are complex-valued generalizations of conditional probability tables.
For a single-qubit gate such as the Hadamard gate in Figure~\ref{fig:noisy_bell_figure}, the conditional amplitude table (Table~\ref{tab:cat_hadamard}) at node \textcolor{blue}{\texttt{q0m1}} will look like the transpose of the $2\times2$ quantum gate unitary matrix.
For quantum gates involving more than one qubit such as the CNOT gate in Figure~\ref{fig:noisy_bell_figure}, the conditional amplitude table (Table~\ref{tab:cat_cnot}) at node \textcolor{orange}{\texttt{q1m3}} will be a permutation of the original quantum gate unitary matrix.
The permutation is possible so long as the unitary matrices have only one non-zero element in each row and column.
This permutation property holds for most elementary quantum gates, and more complex gates can be decomposed until such translation is possible.


\begin{table}[t]
\caption{Conditional amplitude tables for Figure~\ref{fig:noisy_bell_figure}.}
\label{tab:noisy_bell_figure}
\centering
\subfigure[
Conditional amplitude table at node \textcolor{blue}{\texttt{q0m1}} associated with the Hadamard gate (H in Figure~\ref{fig:noisy_bell_figure}).
Table rows list input qubit basis state combinations; table columns list output qubit basis state combinations.
]{
\makebox[.9\linewidth]{
\begin{tabular}{c|cc}
\toprule
\textcolor{green}{\texttt{q0m0}} & $A(\textcolor{blue}{\texttt{q0m1}}=\ket{0})$ & $A(\textcolor{blue}{\texttt{q0m1}}=\ket{1})$\\
\midrule
$\ket{0}$ & $+1/\sqrt{2}$ & $+1/\sqrt{2}$ \\
$\ket{1}$ & $+1/\sqrt{2}$ & $-1/\sqrt{2}$ \\
\bottomrule
\end{tabular}
\label{tab:cat_hadamard}
}
}
\par
\subfigure[
Conditional amplitude table at node \textcolor{black}{\texttt{q0m2rv}} representing probabilities of measurement outcomes for the phase damping noise (PD in Figure~\ref{fig:noisy_bell} and $R_y$ in Figure~\ref{fig:noisy_bell_transformed}).
]{
\makebox[.9\linewidth]{
\begin{tabular}{c|cc}
\toprule
\textcolor{blue}{\texttt{q0m1}} & $A(\textcolor{black}{\texttt{q0m2rv}}=0)$ & $A(\textcolor{black}{\texttt{q0m2rv}}=1)$\\
\midrule
$\ket{0}$ & $1$ & $0$ \\
$\ket{1}$ & $+0.8$ & $-0.6$ \\
\bottomrule
\end{tabular}
\label{tab:cat_phase_damping}
}
}
\par
\subfigure[
Conditional amplitude table at node \textcolor{orange}{\texttt{q1m3}} associated with CNOT gate.
]{
\makebox[.9\linewidth]{
\begin{tabular}{cc|cc}
\toprule
Control \textcolor{blue}{\texttt{q0m1}} & Target \textcolor{red}{\texttt{q1m0}} & $A(\textcolor{orange}{\texttt{q1m3}}=\ket{0})$ & $A(\textcolor{orange}{\texttt{q1m3}}=\ket{1})$\\
\midrule
$\ket{0}$ & $\ket{0}$ & $+1$ & $0$ \\
$\ket{0}$ & $\ket{1}$ & $0$ & $+1$ \\
$\ket{1}$ & $\ket{0}$ & $0$ & $+1$ \\
$\ket{1}$ & $\ket{1}$ & $+1$ & $0$ \\
\bottomrule
\end{tabular}
\label{tab:cat_cnot}
}
}
\end{table}



\subsubsection{Encoding Noisy Quantum Mixtures \& Channels}

In this paper, we propose for the first time additional semantics for representing quantum noise mixtures and channels in quantum Bayesian networks.
For qubit states that follow quantum noise mixtures, the parameters in the conditional amplitude tables come from the probabilities of the noise mixture possibilities and their effect on the quantum state.
For qubit states that follow quantum noise channels, the probability of whether the noise event occurs is encoded in a random variable representing spurious measurement outcomes (Table~\ref{tab:cat_phase_damping}).
Such a representation for quantum noise works for all canonical noise models, including the symmetric and asymmetric depolarizing, bit-flip, phase-flip, (generalized) amplitude damping, and phase damping types of noise listed in Table~\ref{tab:noise_models}.

\begin{table*}[t]
\caption{Program transformations converting Bayesian networks to conjunctive normal form (CNF).}
\label{tab:bn_to_cnf}
\begin{tabularx}{\linewidth}{X|p{5.12cm}p{5.12cm}|X}
\toprule
\textbf{Quantum circuit\newline semantics encoded}
&
\multicolumn{2}{p{10.24cm}|}{\textbf{The interpreted meaning of logical sentences comprising the CNF for our noisy Bell state quantum circuit example in Figure~\ref{fig:noisy_bell_figure} and Table~\ref{tab:noisy_bell_figure}}}
& \textbf{Compilation and \newline simplification rules}
\\
\toprule
Qubits take on binary values; supply known initial qubit values
&
\textcolor{green}{$ \texttt{q0m0}=\ket{0}\;\textsf{XOR}\;\texttt{q0m0}=\ket{1} $}
\newline
\textcolor{green}{$ \texttt{q0m0}=\ket{0} $}
\newline
\textcolor{blue}{$ \texttt{q0m1}=\ket{0}\;\textsf{XOR}\; \texttt{q0m1}=\ket{1} $}
&
\textcolor{red}{$ \texttt{q1m0}=\ket{0} \;\textsf{XOR}\;\texttt{q1m0}=\ket{1} $}
\newline
\textcolor{red}{$ \texttt{q1m0}=\ket{0} $}
\newline
\textcolor{orange}{$ \texttt{q1m3}=\ket{0}\;\textsf{XOR}\; \texttt{q1m3}=\ket{1} $}
&
Combine initial value sentences into binary constraint sentences using logical unit resolution.
\\
\midrule
Hadamard gate (Conditional amplitude table in Table~\ref{tab:cat_hadamard})
&
$ \textcolor{green}{\texttt{q0m0}=\ket{0}} \land \textcolor{blue}{\texttt{q0m1}=\ket{0}} \implies +\frac{1}{\sqrt{2}} $
\newline
$ \textcolor{green}{\texttt{q0m0}=\ket{1}} \land \textcolor{blue}{\texttt{q0m1}=\ket{0}} \implies +\frac{1}{\sqrt{2}} $
\newline
&
$ \textcolor{green}{\texttt{q0m0}=\ket{0}} \land \textcolor{blue}{\texttt{q0m1}=\ket{1}} \implies +\frac{1}{\sqrt{2}} $
\newline
$ \textcolor{green}{\texttt{q0m0}=\ket{1}} \land \textcolor{blue}{\texttt{q0m1}=\ket{1}} \implies -\frac{1}{\sqrt{2}} $ 
&
Compiler needs to avoid simplifications that assume that amplitudes sum to $1.0$.
\\
\midrule
Phase damping noise channel (Conditional amplitude table in Table~\ref{tab:cat_phase_damping})
&
\textcolor{black}{$ \texttt{q0m2rv}=0\;\textsf{XOR}\;\texttt{q0m2rv}=1 $}
\newline
$ \textcolor{blue}{\texttt{q0m1}=\ket{0}} \implies \textcolor{black}{\texttt{q0m2rv}=0} $
&
$ \textcolor{blue}{\texttt{q0m1}=\ket{1}} \land \textcolor{black}{\texttt{q0m2rv}=0} \implies +0.8 $
\newline
$ \textcolor{blue}{\texttt{q0m1}=\ket{1}} \land \textcolor{black}{\texttt{q0m2rv}=1} \implies -0.6 $
&
Weight variables stand in for numerical parameters for amplitudes or probabilities; the simulator later resolves the weight variables with values that can change for repeated simulations.
\\
\midrule
CNOT gate (Conditional amplitude table in Table~\ref{tab:cat_cnot})
&
$ \textcolor{blue}{\texttt{q0m1}=\ket{0}} \land \textcolor{red}{\texttt{q1m0}=\ket{0}} \implies \textcolor{orange}{\texttt{q1m3}=\ket{0}} $
\newline
$ \textcolor{blue}{\texttt{q0m1}=\ket{0}} \land \textcolor{red}{\texttt{q1m0}=\ket{1}} \implies \textcolor{orange}{\texttt{q1m3}=\ket{1}} $
&
$ \textcolor{blue}{\texttt{q0m1}=\ket{1}} \land \textcolor{red}{\texttt{q1m0}=\ket{0}} \implies \textcolor{orange}{\texttt{q1m3}=\ket{1}} $
\newline
$ \textcolor{blue}{\texttt{q0m1}=\ket{1}} \land \textcolor{red}{\texttt{q1m0}=\ket{1}} \implies \textcolor{orange}{\texttt{q1m3}=\ket{0}} $
&
Deterministic parameters such as $0.0$ or $1.0$ can be directly factored into logic without weight variables.
\\
\bottomrule
\end{tabularx}
\end{table*}

\subsection{Bayesian Network Knowledge Compilation}
Now that the semantics of noisy quantum circuits have been compiled into our Bayesian network representation,
we demonstrate for the first time using inference techniques based on logical formula minimization to enable efficient quantum circuit simulation.
There are many algorithms for exact inference on Bayesian networks.
Initially, we used variable elimination~\cite{darwiche2009modeling,Koller:2009:PGM:1795555,Korb:2010:BAI:1941985,pearl2014probabilistic}
to demonstrate that exact inference on the complex-valued Bayesian networks leads to correct circuit simulation results.
We soon realized that support for repeated simulation with different parameters was the key to support important variational algorithms.
The need for repeated inference motivates using exact inference algorithms based on knowledge compilation~\cite{darwiche2002knowledge,Kimmig:2017:AMC:3095823.3095885}.

Knowledge compilation techniques compile Bayesian networks into logical formulas with associated weight values on satisfying sets of variable assignments (Section~\ref{sec:bn_to_cnf}).
Then, these formulas are further compiled into arithmetic circuits that exploit conditional independences in order to minimize their representation, allowing a circuit to be compiled once and queried many times efficiently (Section~\ref{sec:cnf_to_ac}).
A sum-of-products process known as weighted model counting on the compiled representations give exact inference results~\cite{Chavira:2008:PIW:1342435.1342783,darwiche2009modeling}.
In the quantum setting, exact inference supports quantum circuit simulation by determining the amplitudes in the final wavefunction (Section~\ref{sec:sim_samp}).

\subsubsection{Bayesian Networks to Conjunctive Normal Form Logical Formulas}
\label{sec:bn_to_cnf}


The first half of the compilation process is to separate the structural information of the quantum circuit from the amplitude and probability numerical parameters of the circuit.
The compiler does this extraction by converting the Bayesian networks into CNF logical formulas, which in the case of the tools we use are encoded in the standard DIMACS format.\footnote{We found and extended a Bayesian network to CNF compiler originally intended for purely classical probabilities for this stage of our toolchain. \url{https://github.com/gisodal/bayes-to-cnf}.}

Our translation of the meaning of an example CNF is shown in Table~\ref{tab:bn_to_cnf}.
Each of the Boolean variables in the CNF corresponds to either the truth value of some qubit state or an indicator variable for a numerical weight.
The table shows logical sentences that encode information represented in the topology of the quantum circuits and the Bayesian networks.
Some sentences in the CNF represent hard constraints on logical variables and qubit states, such as \textcolor{green}{$ \texttt{q0m0}=\ket{0} $}.
Other sentences encode a weight value assigned to a combination of logical conditions, such as $ \textcolor{blue}{\texttt{q0m1}=\ket{1}} \land \textcolor{black}{\texttt{q0m2rv}=1} \implies -0.6 $.

The resulting CNF from conjoining all the clauses together expresses all the combinations of qubit states that are consistent with the quantum circuit semantics.
Each set of valid variable assignments that satisfies the CNF represents one valid Feynman path~\cite{feynman2006qed} through the quantum circuit.
A weighted model count on the weight values for these satisfying assignments leads to the amplitudes we need to perform quantum circuit simulation.

To our knowledge, ours is the first work to represent and manipulate quantum circuits as logical formulas; such a representation enables us to use logical minimization techniques to aid in circuit simulation.

\paragraph{Optimizations}
The Bayesian network to CNF compiler applies various simplification rules on CNFs at this stage.
\begin{enumerate}
    \item The compiler substitutes known variable values (\textit{e.g.}, known initial qubit states) into other sentences containing the same variable in order to simplify those sentences~\cite{Chavira:2005:EEP:3020336.3020350}.
    \item The compiler recognizes deterministic probabilities such as 0.0 and 1.0 to eliminate irrelevant sentences.
    \item Numerical parameters, such as $-1/\sqrt{2}$ in the Hadamard gate and the 0.36 probability in the phase damping channel, are replaced with variables whose values are resolved later; such a substitution allows the simulator to efficiently repeat simulations with different sets of parameters during simulator execution.
\end{enumerate}
These simplifications lead to a linear reduction in the number of clauses in the CNFs that lead to a significant reduction in later compilation results.

In general, the semantics of translating classical real-valued Bayesian networks to CNFs for knowledge compilation has been the subject of numerous studies~\cite{cd05,Chavira:2006:ECE:2165381.2165393,Chavira:2008:PIW:1342435.1342783,Darwiche:2002:LAF:3087093.3087128,darwiche2009modeling,skb05}.
Some of these optimization techniques assume probabilities that sum to unity~\cite{skb05},
and would therefore lead to an incorrect encoding for quantum simulation on amplitudes.




\begin{figure*}[t]{}
\centering
\includegraphics[width=\linewidth]{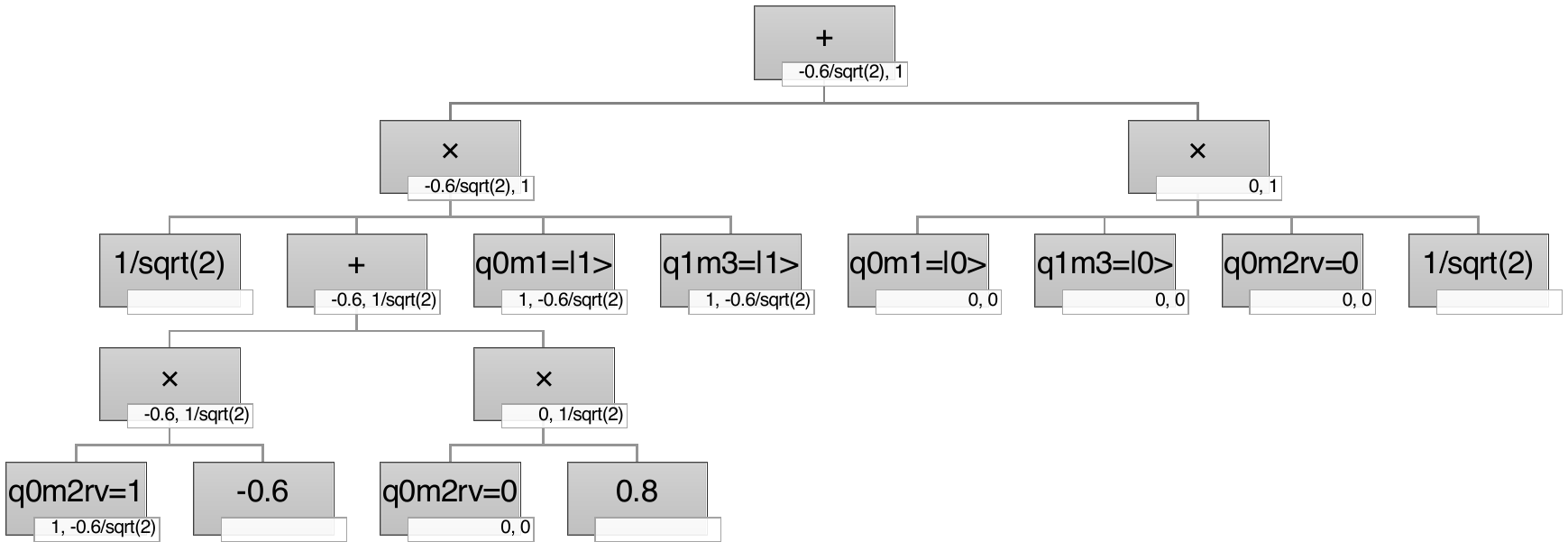}
\caption{The arithmetic circuit resulting from knowledge compilation of the noisy quantum circuit and Bayesian network in Figure~\ref{fig:noisy_bell_figure} and corresponding CNF in Table~\ref{tab:bn_to_cnf}.
This data structure represents combinations of qubit states that are consistent with the quantum circuit semantics, and it assigns a weight to each.
The arithmetic circuit enables efficient amplitude calculation in an upward traversal through the tree (Section~\ref{sec:inference}) and also Gibbs sampling in a downward traversal (Section~\ref{sec:sampling}).
}
\label{fig:arithmetic_circuit}
\end{figure*}

\subsubsection{CNFs to Minimized Arithmetic Circuits}
\label{sec:cnf_to_ac}

The second half of the compilation process is to compile the CNFs into arithmetic circuits (ACs), such as the ones in Figures~\ref{fig:vcg} and~\ref{fig:arithmetic_circuit}, which are data structures that represent the sets of satisfying variable assignments in a minimized representation.
The representation enables calculating the sum of products of the weights for all paths.
In our use of ACs for quantum circuit simulation here, the data structure represents all Feynman paths through the quantum circuit consistent with the given initial qubit states and measurement outcomes, and it allows the simulator to calculate the amplitude for the assigned qubit states for each path.

The tasks of calculating and sampling amplitudes both take place with time complexity linear with respect to the size of the compiled AC~\cite{darwiche2002knowledge,10.5555/2900423.2900456,Kimmig:2017:AMC:3095823.3095885}, so it is worth discussing the costs of such a compiled representation next.

\paragraph{Worst-case complexity of path enumeration and quantum circuit simulation}
The number of satisfying assignments to a CNF grows exponentially in the worst case.
As shown in the "Before" picture in Figure~\ref{fig:vcg}, direct enumeration of satisfying assignments of a CNF representing a quantum circuit leads to ACs that have combinatorially many paths through the circuit.
The power of quantum algorithms arises from the parallel and simultaneous traversal of all edges in the graph, accentuating some basis states in the final wavefunction while cancelling out other basis states.
Such combinatorial explosion in the number of paths is also what makes classical simulation of quantum algorithms intractable.

\paragraph{Complexity of path enumeration in practice with knowledge compilation}
The appeal of the knowledge compilation approach is that various optimizations enable compilation of CNFs to ACs without resulting in exponentially large ACs in practice.
The caveat, however, is that compiling CNFs to minimized ACs may take time exponential with respect to the input CNF size due to the inherent hardness of the factoring and minimization task.
Nonetheless, such a precompilation cost is still worthwhile in simulating variational quantum circuits, where the simulator can reuse the compiled data structure for repeated simulation with different parameters.

\paragraph{Optimizations}
In this compilation stage, various optimization options impact the compiled representations' size.\footnote{We use {\sc c2d} to convert CNFs to ACs.
\url{http://reasoning.cs.ucla.edu/c2d/}. Alternatives such as Dsharp and D4 do not have available all the optimizations useful for this work.}
\begin{enumerate}
    \item \textbf{Qubit state elision.}
Since for the purposes of this paper we are only interested in the final qubit states, we can instruct the compiler to use existential quantification to factor away the variables corresponding to intermediate qubit states.
For example in Figure~\ref{fig:arithmetic_circuit}, the nodes corresponding to the known initial qubit states (\textcolor{green}{\texttt{q0m0}} and \textcolor{red}{\texttt{q1m0}}) and the intermediate qubit state (\textcolor{violet}{\texttt{q0m2}}) are factored out from the compiled AC.
Such elision enables the AC to calculate amplitudes for the output qubits without incurring the unnecessary cost of calculating amplitudes for intermediate qubit states.



    \item \textbf{Qubit state elimination order.}
    The order in which logical variables corresponding to the remaining qubit states are enumerated impacts how much factoring the compiler can perform.
Elimination order choices include using a hypergraph partitioning algorithm and also one that follows the lexicographic ordering for qubit states.
We observe that using hypergraph partitioning allows for smaller AC sizes and therefore faster simulation times when only the final output qubit states are relevant.
\end{enumerate}

These optimizations, in conjunction with the CNF minimization rules in Section~\ref{sec:bn_to_cnf}, lead to a reduction in circuit size demonstrated in the "After" picture in Figure~\ref{fig:vcg}.

\begin{figure}[t]
\centering
\includegraphics[width=\columnwidth]{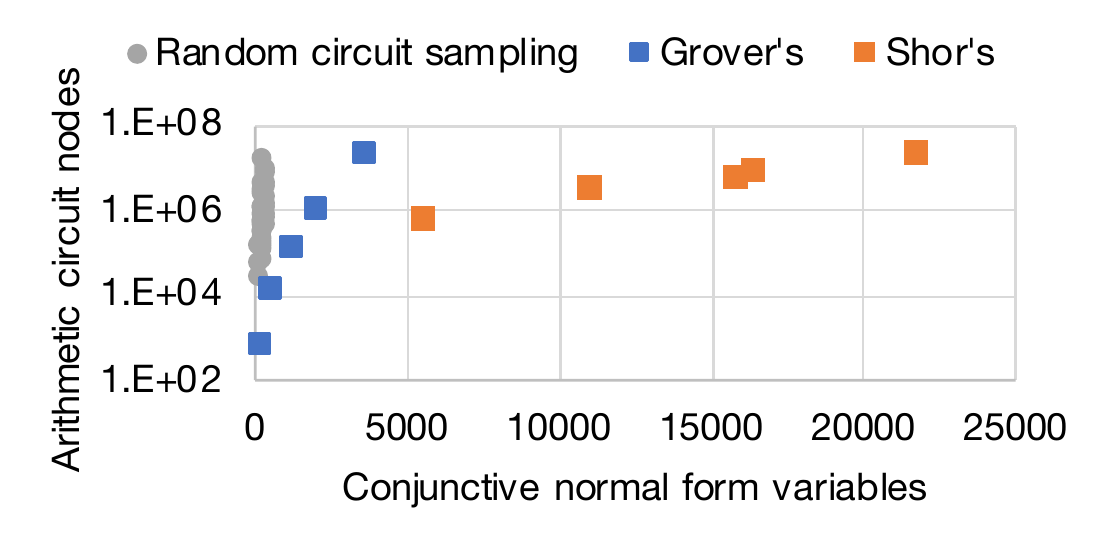}
\caption{
Simulation resource requirements vs. quantum circuit size for three quantum algorithms.
}
\label{fig:resource_requirements}
\end{figure}

\begin{table}[t]
    \caption{Problem size metrics for largest instances in Figure~\ref{fig:resource_requirements}.}
\label{tab:problem_size_metrics}
\centering
    \begin{tabular}{rlll}
      \toprule
      & \# qubits & \# gates & AC file size \\
      \midrule
      RCS & 42 & 840 & 82 MB \\
      Grover's & 17 & 2460 & 530 MB \\
      Shor's & 13 & 12247 & 586 MB \\
      \bottomrule
    \end{tabular}
\end{table}

\subsubsection{Evaluation of Knowledge Compilation on Quantum Algorithm Case Studies}



As we will demonstrate next for a variety of structured and unstructured quantum circuits, the compiled AC representations avoid the worst case and offer reductions in simulation complexity.

Figure~\ref{fig:resource_requirements} plots the resource requirements of a simulation against the underlying quantum circuit size for various quantum algorithm simulation workloads.
The horizontal axis counts the number of variables in the CNF (and in the Bayesian network);
this value is proportional to the number of gates inside the quantum algorithm quantum circuit.
The vertical axis, in log scale, counts the number of nodes in the compiled AC;
this number is proportional to several measures of the simulation resource intensiveness,
including the number of edges in the AC,
the memory and filesize needed to store the AC,
and the time to compile and perform inference on the AC.

The data points in the plot correspond to simulation instances of various sizes
(Table~\ref{tab:problem_size_metrics})
belonging to three quantum algorithms.
Two of the three algorithms are \emph{structured} workloads, meaning that they are designed to perform a meaningful computation. 
The orange data points are instances of Shor's factoring algorithm~\cite{doi:10.1137/S0097539795293172}, written in a style that minimizes qubit count~\cite{beauregard}.
The circuits here are factoring either 6 or 15, covering a range of one through four iterations of the algorithm.
The blue data points are instances of Grover's search algorithm~\cite{Grover:1996:FQM:237814.237866,doi:10.1119/1.1359518}.
In this case the algorithm is searching for the square root of a number in a simple abstract algebra setting, for a search space ranging from two to 16 elements.
The implementations are taken from open source quantum algorithm benchmarks,\footnote{\url{https://github.com/epiqc/ScaffCC}} and the simulation results are validated to be correct outputs.

The third algorithm is an \emph{unstructured} workload, meaning that the quantum operations are randomly selected and placed in a fixed template.
These problems in random circuit sampling (RCS) are extremely difficult to simulate because the qubits rapidly become entangled with all other qubits~\cite{boixo2018supremacy,Harrow,Terhal}, leaving little independence structure for knowledge compilation to exploit.
The gray data points are simulations of a population of such workloads involving between 25 and 42 qubits.\footnote{\url{https://github.com/sboixo/GRCS}}

The trends here on a semi-log plot show different scaling trends among the three workloads.
The RCS workload exhibits full exponential growth in simulation difficulty,
while the Grover's and Shor's workloads appear to scale sub-exponentially.
This is a result of the knowledge compilation toolchain extracting structure with different degrees of success for the three classes of workloads.
The significance of this capability is that we can repurpose knowledge compilation to extract structure and reduce the cost of simulating a quantum circuit.
\subsection{Calculating Amplitudes and Sampling from Arithmetic Circuits}
\label{sec:sim_samp}

The ACs that result from the previous program transformations dictate the minimal sequence of calculations for both finding amplitudes for a given set of qubit states (Section~\ref{sec:inference}) and also sampling outcomes from the final wavefunction (Section~\ref{sec:sampling}), for a given quantum circuit topology and a given variable order.
These two tasks proceed, respectively, as upward and downward traversals of the AC graph.
The ACs memoize calculation results from previous queries so that only changed nodes have to be recalculated for new queries.

ACs such as the one in Figure~\ref{fig:arithmetic_circuit} consists of nodes that are either operations (multiply, add) or leaves.
The leaves represent either numerical parameters---quantum amplitudes (\textit{e.g.}, $1/\sqrt{2}$) and noise probabilities (\textit{e.g.}, $0.6$)---or logical variables representing qubit states (\textit{e.g.}, $\textcolor{blue}{\texttt{q0m1}=\ket{0}}$).
The actual values describing quantum amplitudes and noise probabilities can vary between simulation runs as they vary across variational algorithm iterations.
Likewise, the truth values for the qubit state assignments can vary to find the amplitude of any output qubit state of interest.

\begin{table}[t]
\setlength{\tabcolsep}{4pt}
\caption{Upward pass for finding amplitudes.}
\label{tab:inference}
\centering
  \begin{tabular}{ccc|c|c}
    \toprule
    \textcolor{black}{\texttt{q0m2rv}} &
    \textcolor{blue}{\texttt{q0m1}} &
    \textcolor{orange}{\texttt{q1m3}} &
    amplitude &
    density matrix component\\
    \midrule
    0 & $\ket{0}$ & $\ket{0}$ & $1/\sqrt{2}$ & \multirow{4}{*}{$\begin{bmatrix}+\frac{1}{2}&0&0&+\frac{0.8}{2}\\0&0&0&0\\0&0&0&0\\+\frac{0.8}{2}&0&0&+\frac{0.64}{2}\end{bmatrix}$}\\
    0 & $\ket{0}$ & $\ket{1}$ & $0$ \\
    0 & $\ket{1}$ & $\ket{0}$ & $0$ \\
    0 & $\ket{1}$ & $\ket{1}$ & $0.8/\sqrt{2}$ \\
    \midrule
    1 & $\ket{0}$ & $\ket{0}$ & $0$ & \multirow{4}{*}{$\begin{bmatrix}0&0&0&0\\0&0&0&0\\0&0&0&0\\0&0&0&+\frac{0.36}{2}\end{bmatrix}$}\\
    1 & $\ket{0}$ & $\ket{1}$ & $0$ \\
    1 & $\ket{1}$ & $\ket{0}$ & $0$ \\
    1 & $\ket{1}$ & $\ket{1}$ & $-0.6/\sqrt{2}$ \\
  \bottomrule
\end{tabular}
\end{table}

\subsubsection{Calculating Amplitudes via Inference on ACs}
\label{sec:inference}

Our simulator calculates the amplitude for a given output basis state by finding the probability amplitude of such evidence in the Bayesian network.
Such a calculation proceeds as an upward traversal of the AC in Figure~\ref{fig:arithmetic_circuit} following the procedure by Darwiche~\cite{darwiche_2003,darwiche2009modeling}.

Now, let's see the traversal procedure in action.
The white insets in Figure~\ref{fig:arithmetic_circuit} contain a pair of values: the left one tracks the upward traversal for finding the amplitude while the right one tracks the downward traversal for sampling to be discussed next in Section~\ref{sec:sampling}.
Suppose we want to find the probability amplitude for the $\ket{11}$ output state, given that the \texttt{q0m2rv} noise event does occur.
The simulator assigns the value 1 to the logical variable nodes for $\texttt{q0m2rv}=1$, $\textcolor{blue}{\texttt{q0m1}=\ket{1}}$, and $\textcolor{orange}{\texttt{q1m3}=\ket{1}}$, indicating that they are true; and it assigns 0 to the logical variable nodes that indicate otherwise.
The calculations dictated by the operator nodes lead to a root node value at the top of $-0.6/\sqrt{2}$, corresponding to the probability amplitude of the assigned output qubit state and noise events.

Following the same procedure above, Table~\ref{tab:inference} completes the calculation of the probability amplitude for all other sets of noise event ($\texttt{q0m2rv}$) and qubit states ($\textcolor{blue}{\texttt{q0m1}}$, $\textcolor{orange}{\texttt{q1m3}}$) assignments, leading to probability amplitudes for eight different possibilities.
The two different assignments for the noise event, $\texttt{q0m2rv}=0$ and $\texttt{q0m2rv}=1$, lead to two density matrix components that sum up to the overall density matrix of 
\[\rho=\begin{bmatrix}\frac{1}{2}&0&0&\frac{0.8}{2}\\0&0&0&0\\0&0&0&0\\\frac{0.8}{2}&0&0&\frac{1}{2}\end{bmatrix}
\]
which is exactly the expected final density matrix result in Equation~\ref{eq:final} for the noisy Bell-state creation circuit in Figure~\ref{fig:noisy_bell} with $\ket{00}$ as the input.




It is worth emphasizing that all of the compilation and simulation techniques we describe up to this point are exact and involve no approximations. Because the compilation and simulation is exact, we can validate that our simulator gives exactly the right probability amplitude distributions across all of the measurement outcomes.
We validate the overall simulation approach by creating a new simulator backend for the Google Cirq open-source framework for quantum programming.\footnote{\url{https://github.com/quantumlib/Cirq}}
The simulator passes a suite of randomized validation tests for ideal noise-free state vector simulation and also noisy density matrix simulation.
We also demonstrate correct simulation results for a benchmark suite of quantum algorithms, including the CHSH inequality protocol~\cite{PhysRevLett.23.880}, Deutsch-Jozsa~\cite{1992RSPSA.439..553D}, Bernstein-Vazirani~\cite{doi:10.1137/S0097539796300921}, Simon's~\cite{Simon}, hidden shift~\cite{10.5555/644108.644189}, quantum Fourier transform, 
Shor's~\cite{doi:10.1137/S0097539795293172}, and Grover's~\cite{Grover:1996:FQM:237814.237866} algorithms.



\begin{figure}[t]
\centering
\subfigure[Sampling error for a 16-qubit noise-free QAOA circuit.]{\makebox[\linewidth]{\label{fig:kc_qaoa_accuracy_vs_samples_ideal}\includegraphics[width=\linewidth]{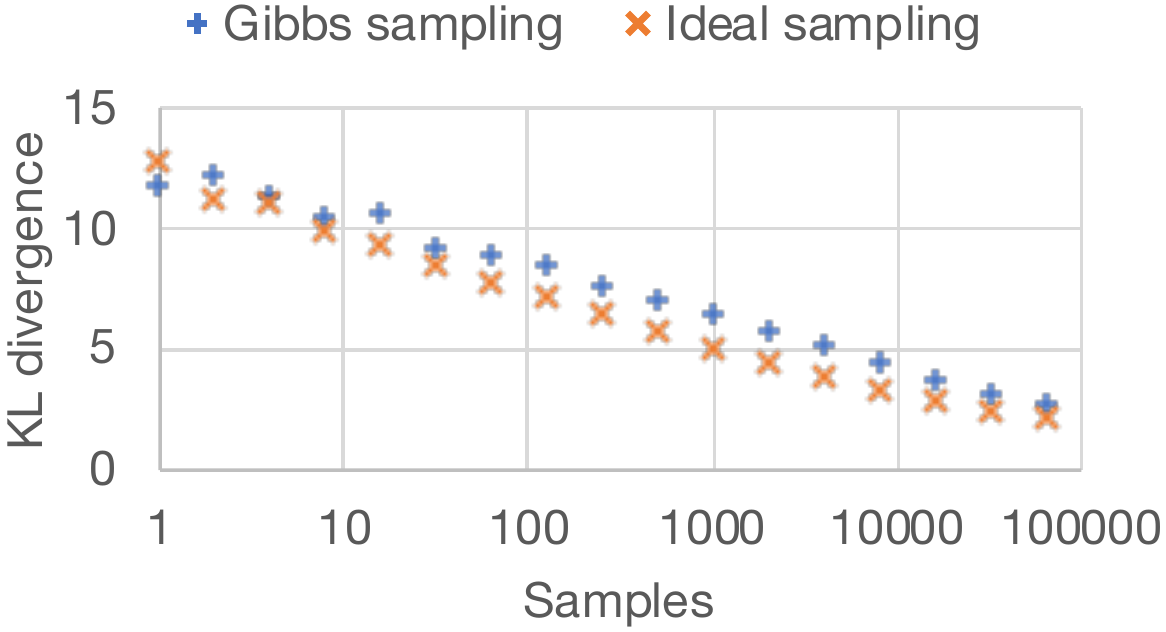}}}
\par
\subfigure[Sampling error for an 8-qubit noisy QAOA circuit.]{\makebox[\linewidth]{\label{fig:kc_qaoa_accuracy_vs_samples_noisy}\includegraphics[width=\linewidth]{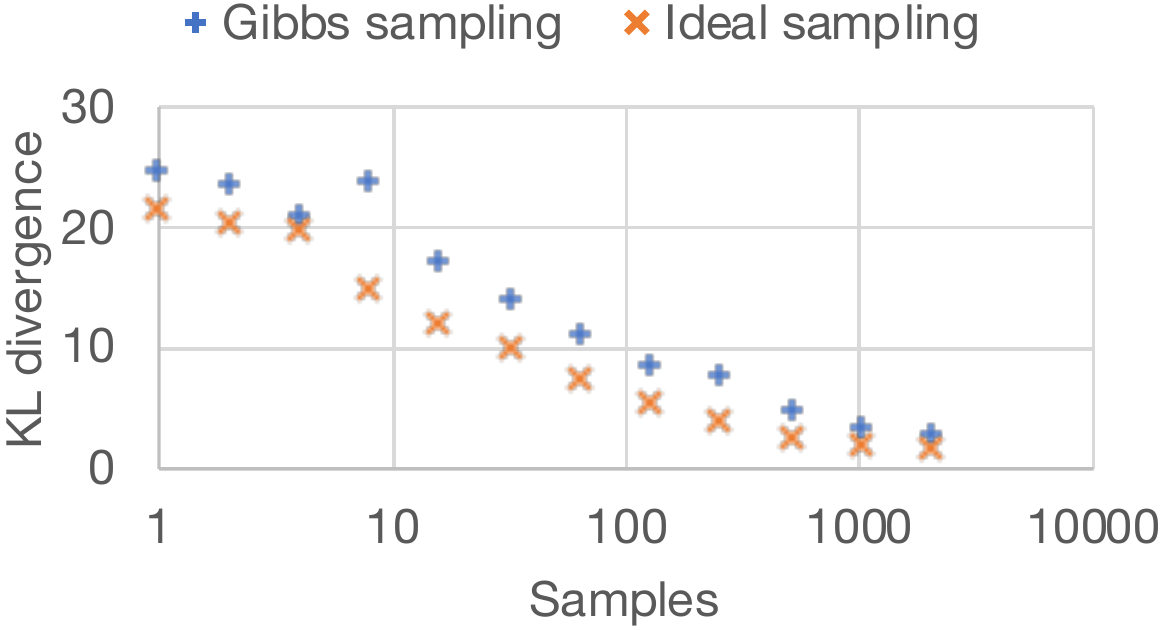}}}
\caption{Sampling error for ideal sampling and Gibbs sampling versus number of samples}
\label{fig:kc_qaoa_accuracy_vs_samples}
\end{figure}

\begin{figure*}[t]
\centering
\subfigure[]{\label{fig:kc_qsim_qaoa_1}\includegraphics[width=.49\linewidth]{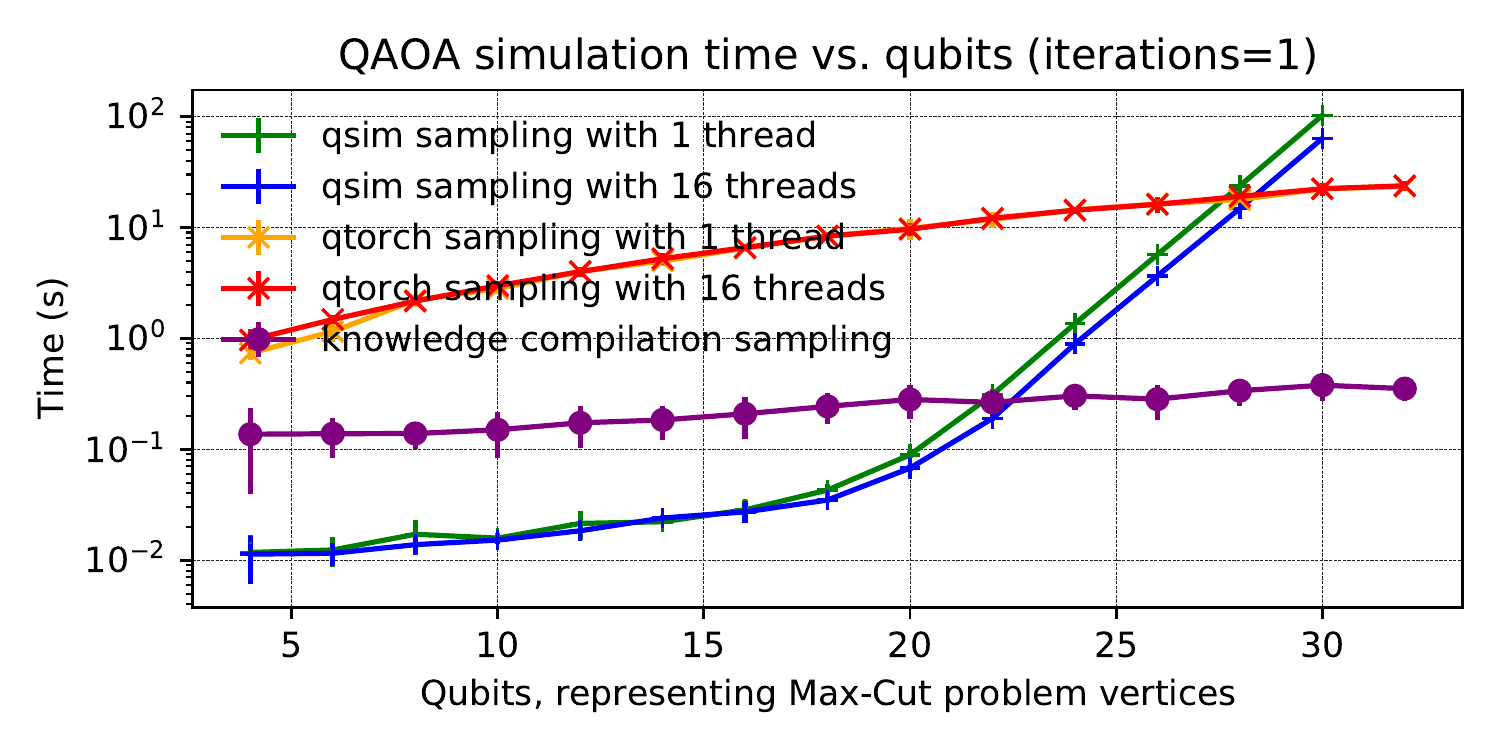}}
\hfill
\subfigure[]{\label{fig:kc_qsim_vqe_1}\includegraphics[width=.49\linewidth]{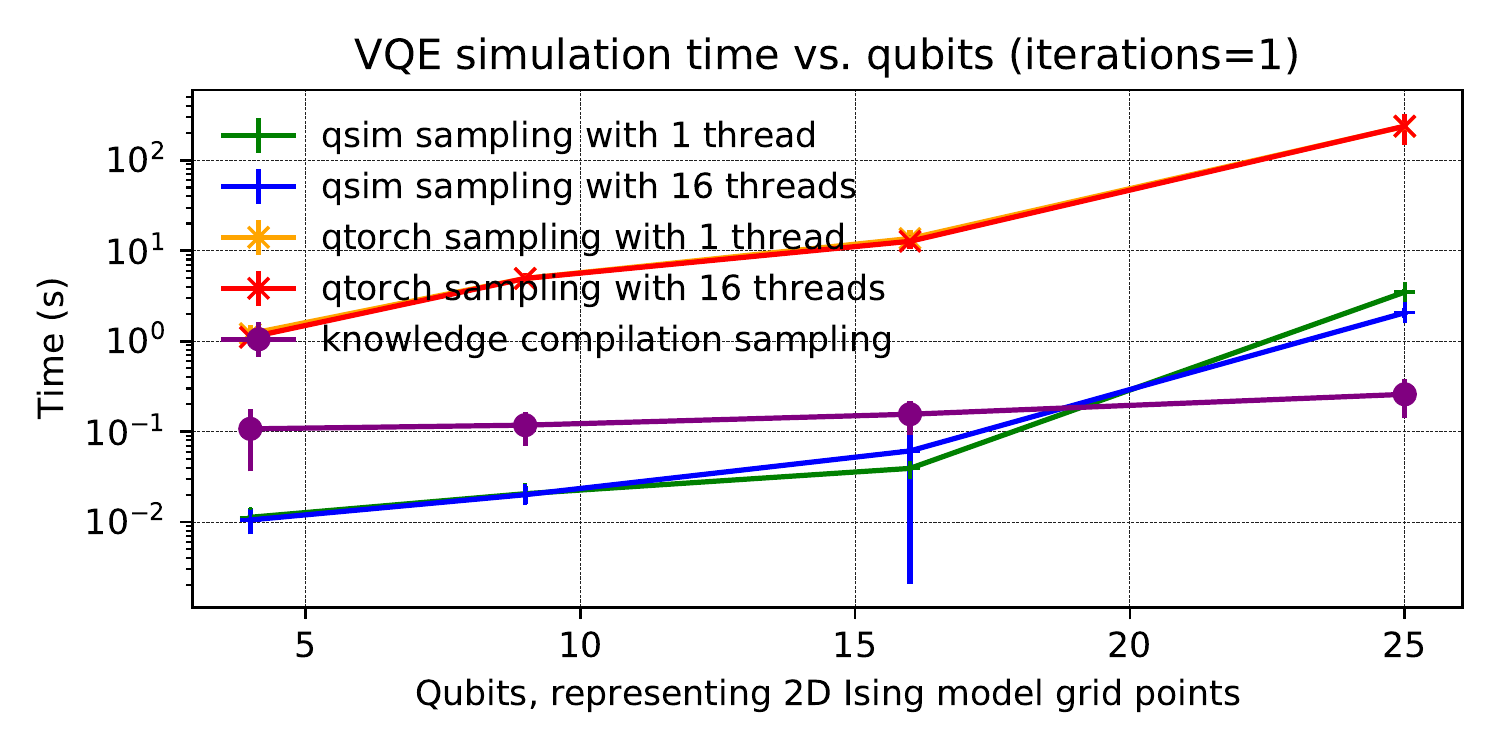}}
\par
\subfigure[]{\label{fig:kc_qsim_qaoa_2}\includegraphics[width=.49\linewidth]{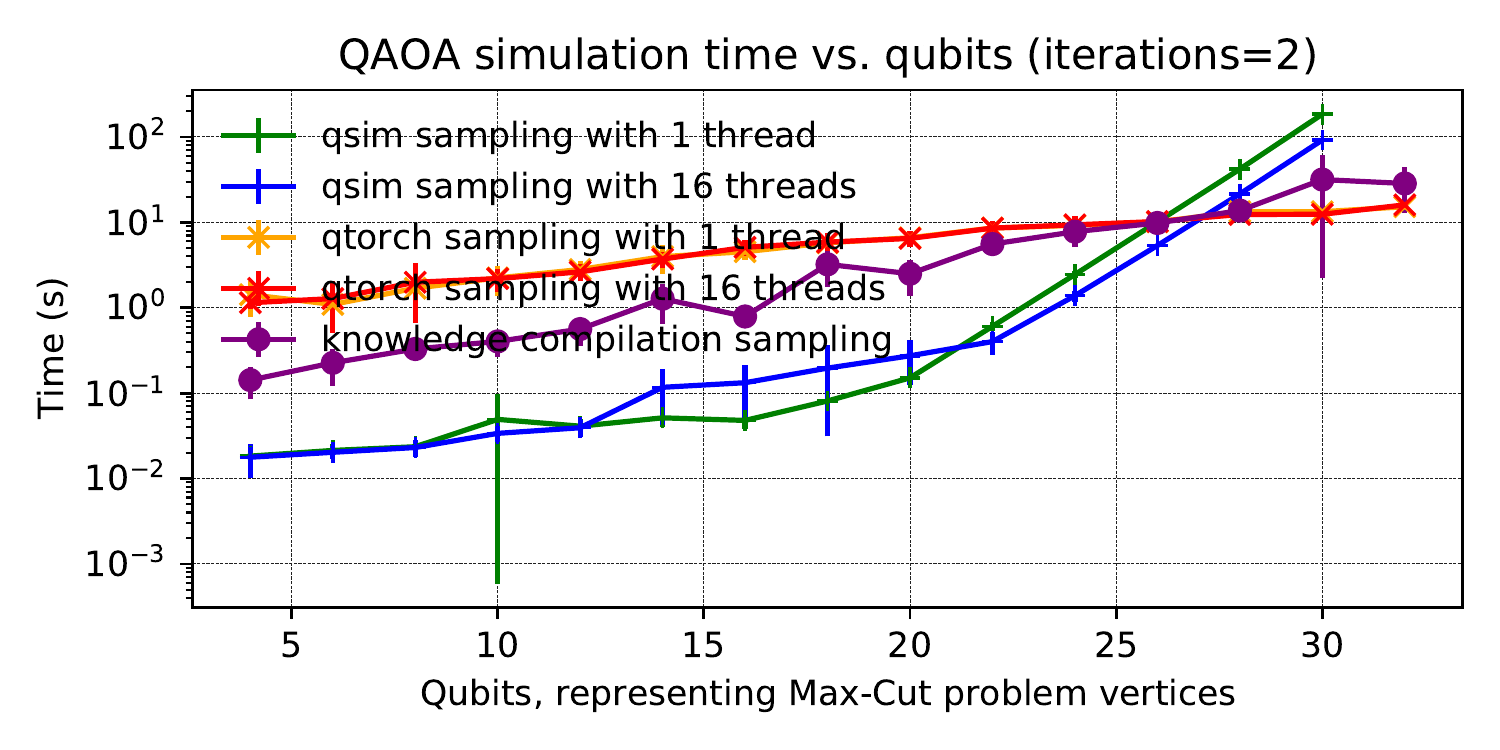}}
\hfill
\subfigure[]{\label{fig:kc_qsim_vqe_2}\includegraphics[width=.49\linewidth]{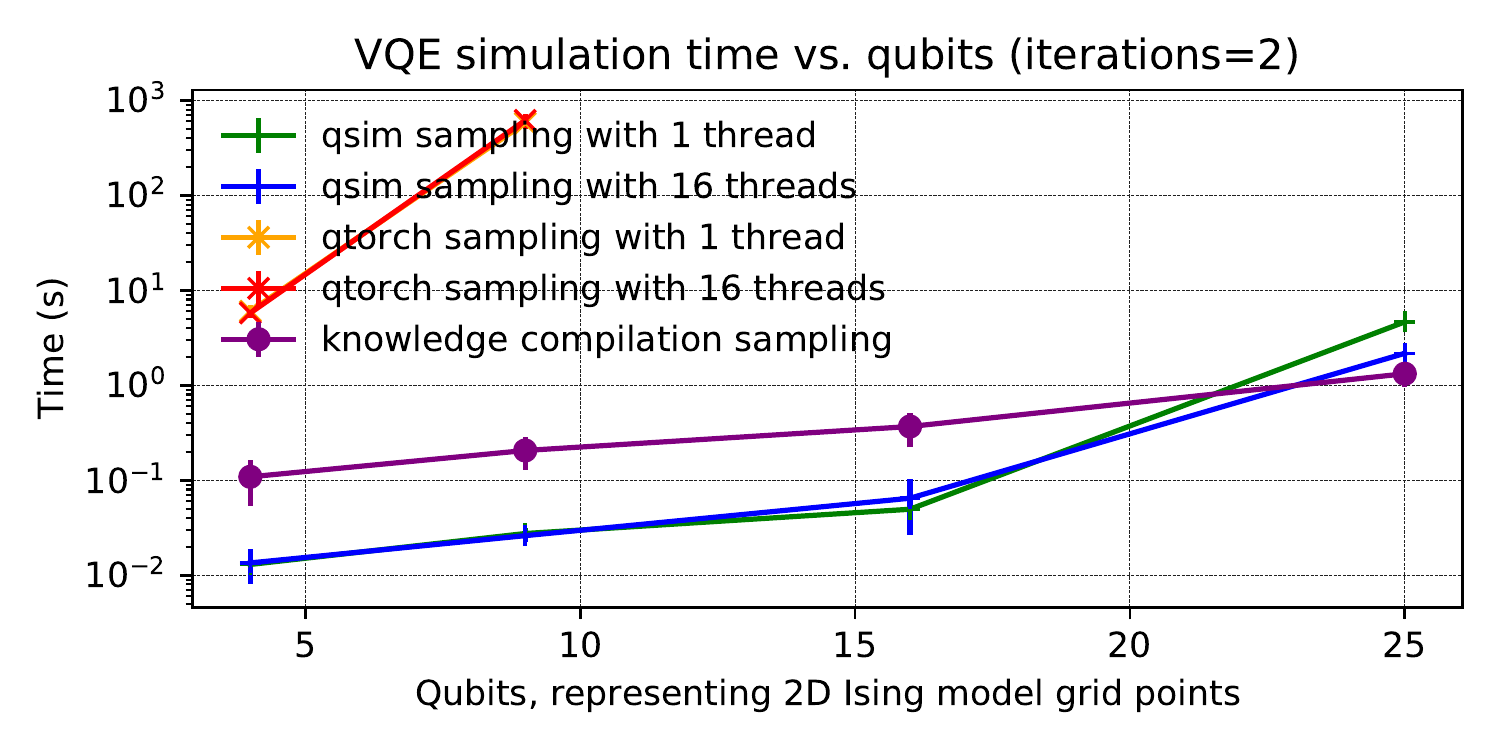}}
\caption{Knowledge compilation vs. state vector (qsim) vs. tensor network (qTorch) performance for ideal (noise-free) circuits}
\label{fig:ideal}
\end{figure*}

\subsubsection{Drawing Samples with Distributions Matching the Output Wavefunction}
\label{sec:sampling}

The final step of our toolchain is to use the compiled arithmetic circuits to approximately sample from the final wavefunction.
Such a feature is important in simulating variational quantum algorithms where a few high-probability quantum measurement outcomes most strongly influence the classical optimizer's objective function (Figure~\ref{fig:noise_qaoa_1024}).
Since a few basis states dominate the output wavefunction, it is easier for the simulator to sample those high probability outcomes than to calculate the full wavefunction.
This task of obtaining a sequence of outcomes matching the probability distribution from measuring the final wavefunction amounts to a Markov chain Monte Carlo (MCMC) problem~\cite{chib1995understanding}.




The compiled AC representation of noisy quantum circuits facilitates Gibbs sampling.
Gibbs sampling is a form of MCMC where the next sample of variable assignments is drawn from the variable assignments that are “one away” from the current assignment.
More concretely, if the present variable assignment is \{ $\texttt{q0m2rv}=1$, $\textcolor{blue}{\texttt{q0m1}=\ket{1}}$, $\textcolor{orange}{\texttt{q1m3}=\ket{1}}$ \}, then the Gibbs sampling MC would consider the following possibilities as the next sample:
\vfill\null
\begin{itemize}
    \item \{ $\texttt{q0m2rv}=0$, $\textcolor{blue}{\texttt{q0m1}=\ket{1}}$, $\textcolor{orange}{\texttt{q1m3}=\ket{1}}$ \}
    \item \{ $\texttt{q0m2rv}=1$, $\textcolor{blue}{\texttt{q0m1}=\ket{0}}$, $\textcolor{orange}{\texttt{q1m3}=\ket{1}}$ \}
    \item \{ $\texttt{q0m2rv}=1$, $\textcolor{blue}{\texttt{q0m1}=\ket{1}}$, $\textcolor{orange}{\texttt{q1m3}=\ket{0}}$ \}
\end{itemize}

Each of these assignments has one of the variable assignments flipped with respect to the present sample.

The compiled arithmetic circuit offers a way to compute the probability of each of the transitions as a downward traversal of the AC~\cite{darwiche2009modeling}.
To find the transition probabilities, we first find the amplitude for a given set of parameters and evidence via an upward traversal in the AC as in Section~\ref{sec:inference}, filling in the left number in the white insets in Figure~\ref{fig:arithmetic_circuit}.
Then, in a downward traversal of the AC, we add to the right number of each node the node's contribution to the final amplitude, following the procedure by Darwiche~\cite{darwiche_2003}.
This right number is the transition probability that helps pick the next step in the MCMC chain.

\subsubsection{Evaluation of Sampling Accuracy for Noise-Free and Noisy Circuits}

Now, we quantify the extent to which the Gibbs MCMC sampling technique facilitated by arithmetic circuits returns the same distribution, compared to ideal (direct) sampling from a fully-known final wavefunction.
We are interested in this evaluation because this final sampling step in our simulation toolchain is an approximation technique, in contrast to the prior steps where the noisy quantum circuits are \emph{exactly} translated into Bayesian networks, CNFs, and arithmetic circuits with no approximation.
As shown in Figure~\ref{fig:noise_qaoa_1024}, the distribution given by Gibbs sampling (Figure~\ref{fig:noise_qaoa_1024}d) introduces inaccuracy compared to ideal sampling (Figure~\ref{fig:noise_qaoa_1024}c) from a known wavefunction (Figures~\ref{fig:noise_qaoa_1024}a and~\ref{fig:noise_qaoa_1024}b).
This inaccuracy is due to warmup and mixing requirements for the Gibbs sampling MCMC.

Figure~\ref{fig:kc_qaoa_accuracy_vs_samples} plots the error of Gibbs sampling and ideal sampling versus the number of samples taken, for simulating both a noise-free and a noisy quantum circuit.
We use the Kullback-Leibler divergence (relative entropy) metric~\cite[Chapter 2.8]{10.5555/971143} to quantify the difference between the sampled distribution versus the fully known distribution.
We choose this metric (in contrast to other metrics such as $\chi^2$) because the KL divergence discounts any error due to zero samples being drawn from low-probability outcomes.
We sample outcomes from a QAOA Max-Cut benchmark circuit involving 16 qubits in the noise-free and 8 qubits in the noisy case.
In the noisy case, the noise model is a symmetric depolarizing noise channel with 0.5\% probability of occurence after each gate.
The trends show that both sampling approaches converge to the same distribution with increasing number of samples.
The Gibbs sampling approach has slightly worse accuracy versus ideal sampling due to the aforementioned MCMC warmup and mixing issues.

This evaluation shows that, for simulating variational algorithms where measurement probabilities are sharply peaked,
the Gibbs sampling approach facilitated via knowledge compilation returns a correct distribution with sufficient samples.

\begin{figure*}[t]
\centering
\subfigure[]{\label{fig:kc_qsim_noise_qaoa_1}\includegraphics[width=0.49\linewidth]{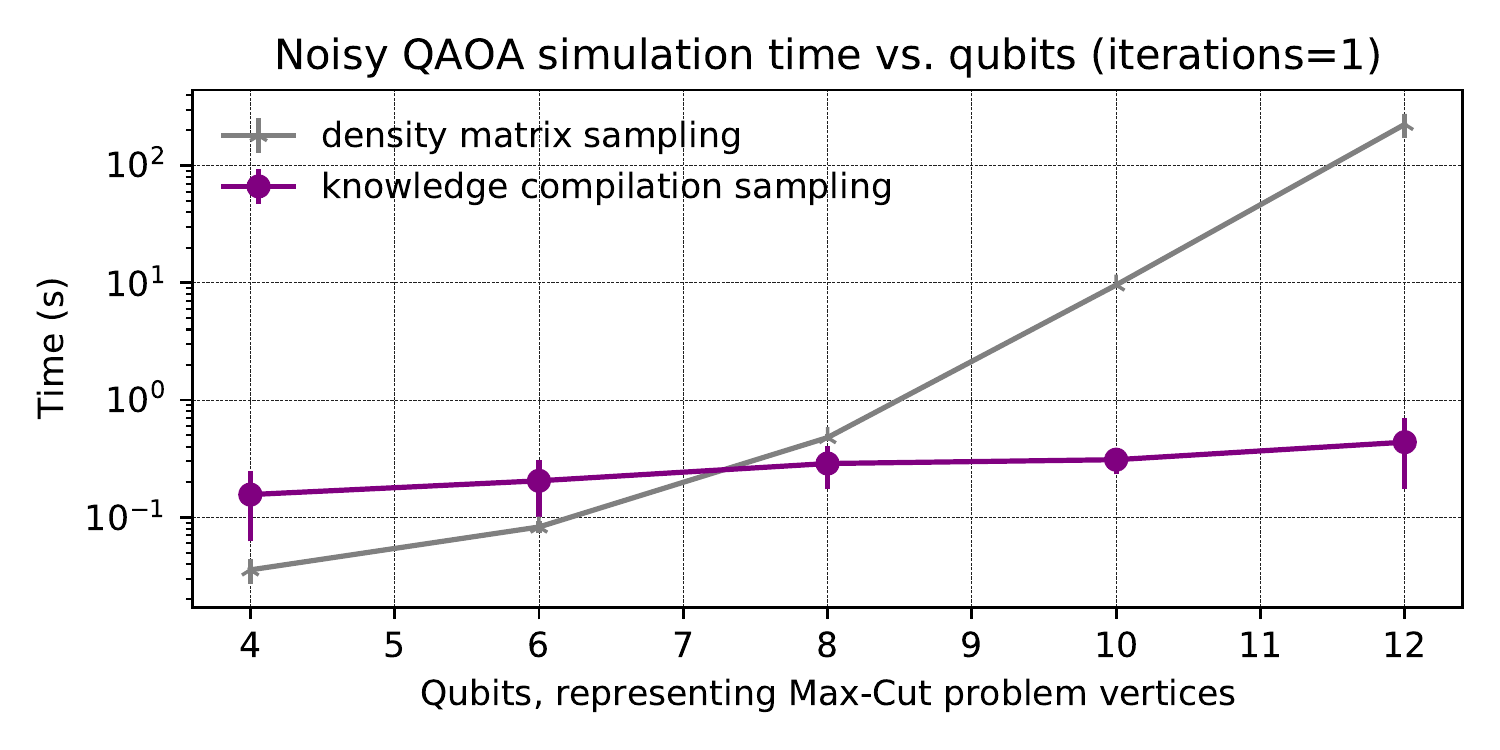}}
\hfill
\subfigure[]{\label{fig:kc_qsim_noise_vqe_1}\includegraphics[width=0.49\linewidth]{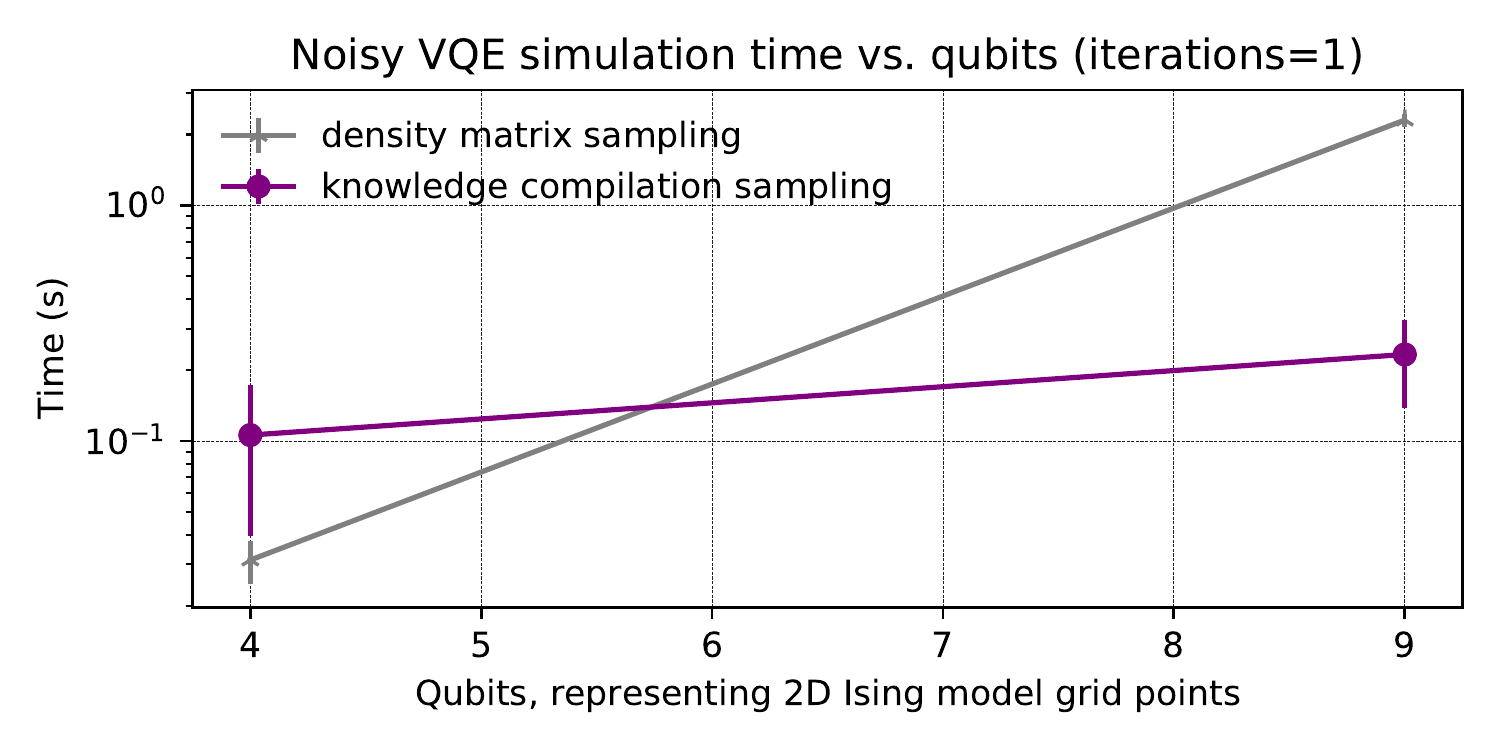}}
\par
\subfigure[]{\label{fig:kc_qsim_noise_qaoa_2}\includegraphics[width=0.49\linewidth]{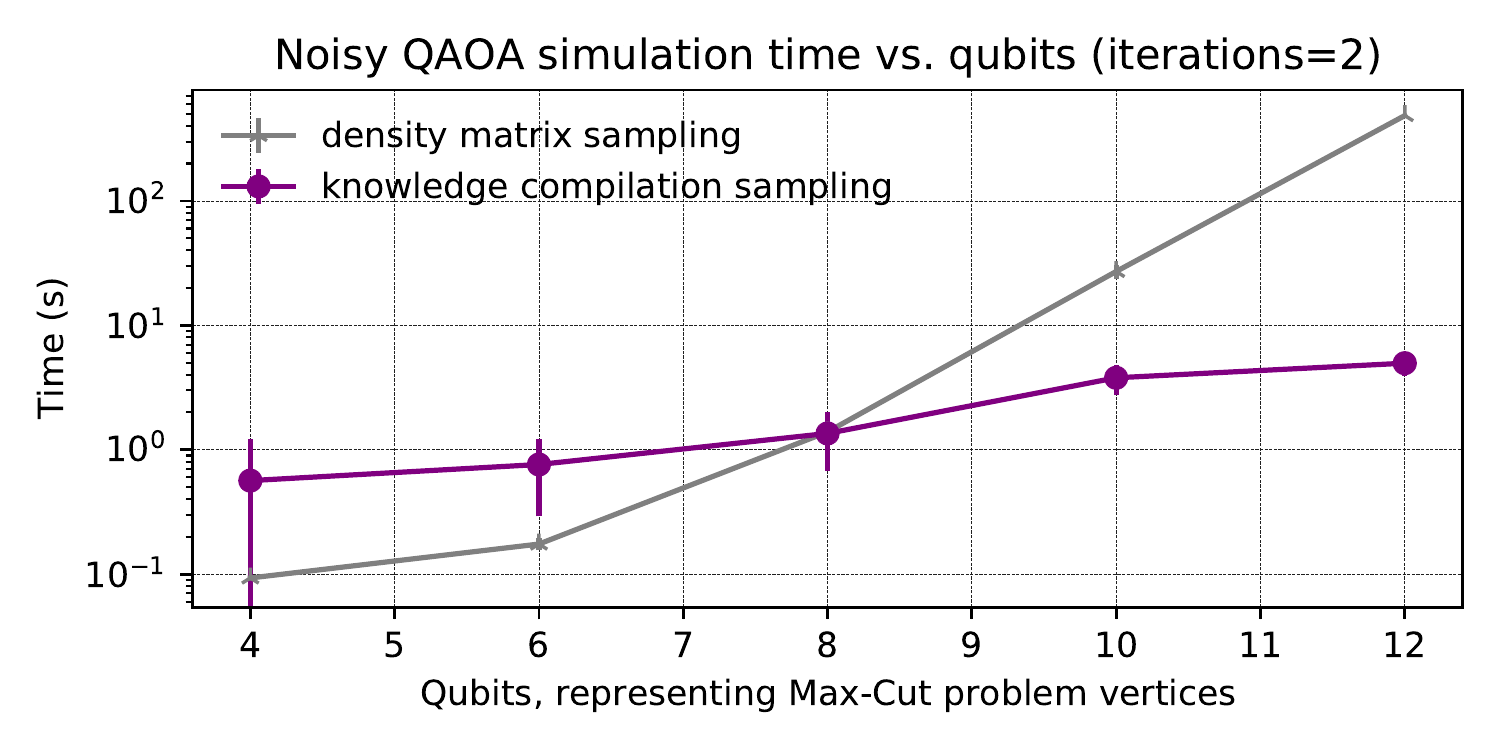}}
\hfill
\subfigure[]{\label{fig:kc_qsim_noise_vqe_2}\includegraphics[width=0.49\linewidth]{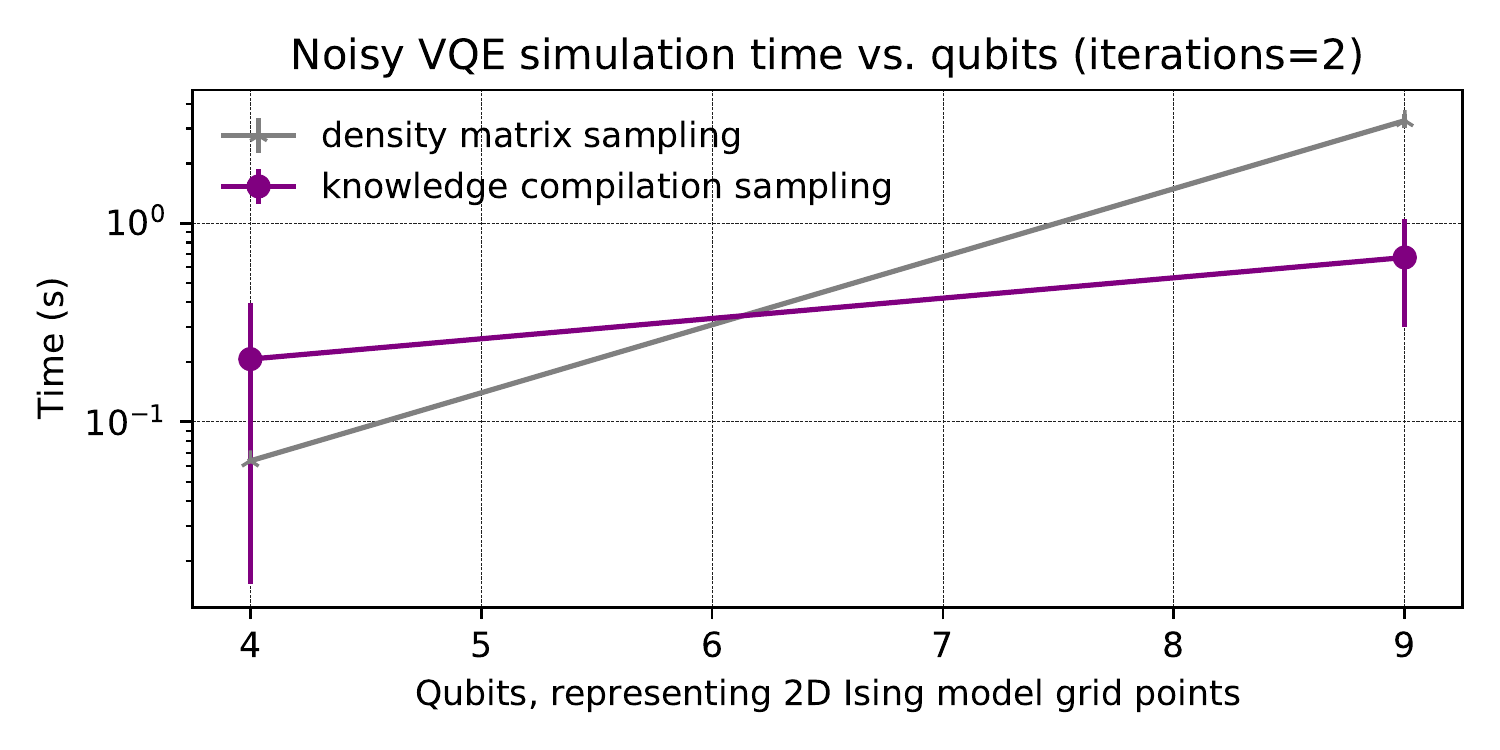}}
\caption{Knowledge compilation vs. density matrix simulation performance for noisy circuits}
\label{fig:noisy}
\end{figure*}

\begin{table*}[t]
    \caption{Intermediate compilation result metrics for largest problem instances in Figures~\ref{fig:ideal} and~\ref{fig:noisy}.}
\label{tab:variational_problem_size_metrics}
\centering
    \begin{tabularx}{\linewidth}{crXXXXXX}
      \toprule
      && \# qubits & \# gates\newline(\# BN nodes) & \# CNF clauses & \# AC nodes & \# AC edges & AC size \\
      \midrule
      & QAOA 1 iteration & 32 & 240 & 1440 & 3139 & 7959 & 50.7 KB \\
      Ideal & QAOA 2 iterations & 32 & 416 & 2592 & 3934271 & 9635580 & 84.8 MB \\
      (Figure~\ref{fig:ideal}) & VQE 1 iteration & 25 & 169 & 839 & 1334 & 2975 & 26.2 KB \\
      & VQE 2 iterations & 25 & 309 & 1579 & 79056 & 188328 & 1.4 MB \\
      \midrule
      & QAOA 1 iteration & 12 & 378 & 3996 & 5798 & 9415 & 43 KB \\
      Noisy & QAOA 2 iterations & 12 & 516 & 5292 & 18991 & 39839 & 304.4 KB \\
      (Figure~\ref{fig:noisy}) & VQE 1 iteration & 9 & 272 & 2867 & 3810 & 5758 & 47.9 KB \\
      & VQE 2 iterations & 9 & 343 & 3637 & 7701 & 13941 & 107.5 KB \\
      \bottomrule
    \end{tabularx}
\end{table*}

\section{Evaluation of Sampling Performance for Ideal and Noisy Circuits}
\label{sec:evaluation}
With the correctness of our knowledge compilation and simulation approach established,
in this section we benchmark our approach on variational algorithm noisy quantum circuits.
We compare against three existing major classes of quantum circuit simulators: state vector, density matrix, and tensor network based simulators.

For problem sizes corresponding to near-term quantum applications (beyond eight noisy qubits with \textasciitilde12 gates per qubit), we demonstrate that the knowledge compilation approach has an advantage over the simulators that tabulate the entire quantum state (\emph{i.e.}, state vector and density matrix simulators), while knowledge compilation's performance advantage relative to tensor network methods depends on the circuit topology.

\subsection{Evaluation for Ideal Circuit Simulation and Sampling}

In Figure~\ref{fig:ideal}, we compare our simulator against qsim,\footnote{\url{https://github.com/quantumlib/qsim}} a state vector simulator by Google that was a component of their quantum supremacy validation experiments~\cite{quantum_ai_team_and_collaborators_2020_4023103}.
qsim is a C-based SIMD simulator that works by multiplying gate unitary matrices against a large state vector.
We also compare against qTorch,\footnote{\url{https://github.com/aspuru-guzik-group/qtorch}} a tensor-network simulator~\cite{2017arXiv170903636S}.
We selected qTorch as a comparison baseline because it is recent, open source, and intended for arbitrary quantum circuits, in contrast to other tensor network simulators that have stipulations on qubit connectivity and the type of simulated quantum circuit.

We evaluate the simulators on two representative variational algorithms, QAOA and VQE.
The QAOA workload (Figure~\ref{fig:kc_qsim_qaoa_1} and~\ref{fig:kc_qsim_qaoa_2}) solves a Max-Cut problem on random graphs with varying number of vertices each having three edges.
Each qubit encodes one vertex, and two-qubit gates between qubits encode the connectivity of the random graphs~\cite{farhi_qaoa,szegedy2019qaoa}.
The VQE workload (Figure~\ref{fig:kc_qsim_vqe_1} and~\ref{fig:kc_qsim_vqe_2}) finds the minimum energy configuration for a 2D Ising model problem.
Each qubit encodes a grid point in 2D space, and two-qubit gates between qubits encode couplings between electron spins~\cite{Barahona_1982}.
For both problems, we perform one or two iterations of the quantum circuit, where the two iteration version would have higher concentration of higher probability outcomes, at the cost of doubling the circuit depth.

For each problem combination, we plot the time it takes to draw 1000 samples against the number of qubits.
The data points are averages across \textasciitilde16 Nelder-Mead optimization runs with randomized problem instances.

At 30 qubits, the qsim state vector simulator has to hold in memory a vector of $2^{30}\approx1.1\mathrm{B}$ complex numbers, which accounts for state vector simulation's exponential cost per simulation run relative to the number of qubits.
The knowledge compilation and tensor network simulators use circuit representations that avoid such a storage cost.
Table~\ref{tab:variational_problem_size_metrics} summarizes metrics for the knowledge compilation intermediate results for the largest problem instances.

The ability for knowledge compilation and tensor network simulators to handle circuit depth (in the form of algorithm iterations) depends on the quantum circuit topology:
At one algorithm iteration, knowledge compilation needs $66\times$ less time than the tensor network method per sample for 32-qubit QAOA;
at two algorithm iterations, the two approaches are comparable for QAOA while qTorch struggles for VQE~\cite{2017arXiv170903636S}.


The results show that for wide (more than 20 noise-free qubits) and shallow (\textasciitilde12 gates per qubit) circuits, the knowledge compilation approach excels at drawing samples from the output wavefunction.
These time savings accumulate over the course of a full simulation for a variational quantum algorithm, as the classical optimizer would draw from these distributions many times in order to evaluate the objective function for different input parameters.


\subsection{Evaluation for Noisy Circuit Simulation and Sampling}

In Figure~\ref{fig:noisy}, we compare our simulator against the density matrix simulator for noisy circuits in Google Cirq.
The density matrix simulator is a NumPy-based simulator that works by multiplying gate unitary matrices against a large density matrix for mixed quantum states.

We evaluate on QAOA and VQE as before, this time adding a symmetric depolarizing noise channel with 0.5\% probability that one of Pauli-X, Y, or Z noise events may happen after each gate.
We further validate that the knowledge compilation simulator calculates the same density matrix as the baseline Google Cirq simulator.

At 12 qubits, the Google Cirq density matrix simulator has to hold in memory a matrix of $2^{12}\times2^{12}\approx17M$ complex numbers; furthermore the matrix has little sparsity to reduce its representation.
Table~\ref{tab:variational_problem_size_metrics} again summarizes metrics for the knowledge compilation intermediate results.

For noisy circuits, the knowledge compilation approach breaks even with the density matrix simulator at eight qubits, fewer than the case for ideal circuits.
This is due to the even greater cost of having to perform matrix-matrix multiplication in density matrix simulation, and also due to less prior focus in developing high performance simulators such as qsim and qTorch for noisy circuit simulation.
The data suggest that knowledge compilation is well-suited for the repeated simulation of noisy circuits in variational quantum algorithms.








\begin{table}[t]
\small
\caption{Comparison of quantum and probabilistic graphical models of computation.}
\label{tab:models_of_computation}
\centering
\begin{tabularx}{\linewidth}{p{0.1cm} X X}
\toprule
 & \textbf{Probabilistic} & \textbf{Quantum} \\\midrule
\parbox[t]{2mm}{\multirow{8}{*}{\rotatebox[origin=c]{90}{Key analogies}}}
 & inference & program simulation \\
 & random variables & qubits \\
 & probabilities & amplitudes \\
 & conditional probability tables & operator unitary matrices \\
 & joint probability distributions & superposition states \\
 & dependent random variables & entangled qubits \\
 & variable elimination & tensor network contraction \\
 & weighted model counting~\cite{Kimmig:2017:AMC:3095823.3095885} & Feynman path sum \\
 \midrule
\parbox[t]{2mm}{\multirow{4}{*}{\rotatebox[origin=c]{90}{Key distinctions}}}
 & probabilities between 0 and 1 & amplitudes are complex-valued \\
 & probabilities sum to 1 & squares of absolute amplitudes sum to 1 \\
 & interference impossible & interference (canceling of amplitudes) possible \\
 & equivalent to Clifford set~\cite{VanDenNes:2010:CSQ:2011350.2011356} & beyond Clifford gate set \\
\bottomrule
\end{tabularx}
\end{table}

\section{Research Directions}
\label{sec:future}

Compiling noisy quantum circuits to PGMs and logical abstractions such as CNFs and arithmetic circuits may accelerate the pace of quantum computing research by offering new ways to analyze quantum circuits.
Table~\ref{tab:models_of_computation} lays out loosely analogous concepts between probabilistic and quantum graphical models.
The fact that Bayesian networks can be generalized to work on complex-valued quantum amplitudes~\cite{tucci1995quantum},
along with the insight that knowledge compilation works on algebraic semirings such as complex numbers~\cite{Kimmig:2017:AMC:3095823.3095885},
underpins the validity of our quantum circuit compilation and simulation toolchain.

This work has focused on finding amplitudes for qubit state assignments and sampling from wavefunctions, by performing a procedure analogous to finding evidence probabilities and gradients via weighted model counting in classical PGMs.
The fact that our quantum circuit simulator gives correct results suggests that other types of PGM query techniques~\cite{HoltzenUAI19,HoltzenOOPSLA20} can likewise support quantum computing research.

Bayesian networks support various other queries such as sensitivity analysis~\cite[Chapter~16]{darwiche2009modeling}\cite{384252} and most probable explanation (MPE) queries.
Sensitivity analysis queries would answer how internal qubit states influence observed qubit states, which may have applications in mapping the most influential qubits variables in an algorithm to the most reliable hardware qubits in a prototype quantum computer.
MPE queries would answer what error event best explains a given symptomatic observed outcome. MPE queries rely on the existence of a meaningful operator for finding the maximum value of two quantities; while such a MAX operator is undefined for complex-valued amplitudes, it does exist for real-valued error probabilities.
These other types of queries can be made tractable, depending on the algebraic properties of what the Bayesian networks represent, and depending on the choice of the knowledge compilation target representation~\cite{10.5555/2900423.2900456,Kimmig:2017:AMC:3095823.3095885,ProbCirc20}.

\section{Conclusion}
This paper proposes and evaluates a new quantum circuit simulation technique that focuses on simulating NISQ era variational quantum algorithms.
Our simulation toolchain extends techniques originating in classical exact probabilistic inference to support this important quantum simulation workload.
Our simulator compiles noisy quantum circuits into complex-valued Bayesian networks in order to combine real-valued noise probabilities and complex-valued quantum amplitudes in one graphical notation.
Our simulator then uses knowledge compilation, a technique originally meant for repeated inference, to form an arithmetic circuit that encodes structure information about the quantum circuit.
The pre-compiled information then allows for efficient repeated quantum circuit simulation with different parameters, and allows for efficient Gibbs sampling from the output wavefunction.
We validated the simulation approach for a benchmark suite of quantum algorithms.
For wide and shallow quantum circuits found in variational algorithms such as QAOA and VQE, our simulator performance compares favorably against both ideal and noisy quantum circuit simulators.
These simulation capabilities may accelerate the development of useful quantum computing systems and near-term quantum algorithms.

\begin{acks}
This work is funded in part by EPiQC,
an NSF Expedition in Computing,
under grant 1730082.
This work is supported in part by NSF grants \#CCF-1837129, \#IIS-1956441, \#IIS-1943641, DARPA grant \#N66001-17-2-4032, a Sloan Fellowship, and gifts from Intel and Facebook Research.
\end{acks}

\appendix
\lstset{basicstyle=\ttfamily,
  showstringspaces=false,
  commentstyle=\color{red},
  keywordstyle=\color{blue}
}

\section{Artifact Appendix}

\subsection{Abstract}

This artifact demonstrates a new way to perform quantum circuit simulation.
We convert quantum circuits into probabilistic graphical models, which are then compiled into a format that enables efficient repeated queries.

The artifact consists of a Docker image which includes Google Cirq, a quantum programming framework, which we have extended to use our proposed approach as a quantum circuit simulation backend.
Also in the Docker image are two quantum circuit simulators based on existing approaches which we compare against as evaluation baselines.

We offer the Docker image via three routes: a hosted version on Docker Hub provides the latest version of our software and requires minimal setup; a Dockerfile is provided to show how to replicate our environment from scratch; and finally a stable archival version is available on Zenodo.

With minimal setup, you can run test cases in our Docker container showing the validity of our approach.
We test our quantum circuit simulation approach using the randomized test harness that Google Cirq uses to test its quantum circuit simulation back ends.
We also demonstrate correct simulation results for a benchmark suite of quantum algorithms.

The Docker image contains performance benchmarking experiments that replicate results of our paper at reduced input problem sizes.
The experiment scripts generate PDFs showing graphs that plot simulation wall clock time against input quantum circuit sizes.
The input problem sizes are large enough to show that our proposed approach achieves a speedup versus existing simulation tools.

\subsection{Artifact Check-List (Meta-Information)}

{\small
\begin{itemize}
  \item {\bf Algorithm: } A new algorithm for simulating quantum circuits and quantum noise models.
  \item {\bf Program: } Google Cirq \url{https://github.com/quantumlib/Cirq}, UCLA Ace compiler \url{http://reasoning.cs.ucla.edu/ace/}, Google qsim \url{https://github.com/quantumlib/qsim}, qTorch \url{https://github.com/aspuru-guzik-group/qtorch}.
  \item {\bf Transformations: } The quantum circuits and noise models are converted to complex-valued Bayesian networks.
  A set of techniques originating in Bayesian inference, known as knowledge compilation, converts the Bayesian networks into logical formulas that support repeated queries.
  \item {\bf Model: } A benchmark suite of quantum circuits provided in Google Cirq.
  \item {\bf Run-time environment: } The Docker container has been tested on Linux (Ubuntu 18.04.5 LTS) and macOS (Big Sur Version 11.1).
  \item {\bf Hardware: } 8 GB memory is needed to run the reduced-size validation test suites. Additional memory (up to 1 TB) is recommended to replicate the paper results for the largest problem instances.
  \item {\bf Execution: } Less than 30 minutes to run the reduced size validation test suites.
  \item {\bf Metrics: } Quantum circuit simulation times for our proposed simulator compared against three baseline simulators from prior work.
  \item {\bf Output: } Four PDF files plotting wall clock times for sampling outputs plotted against quantum circuit size.
  \item {\bf Experiments: } Pull Docker image (or load from tarball), run Docker container, and call various Python scripts within container.
  \item {\bf How much disk space required (approximately)?: } 4 GB
  \item {\bf How much time is needed to prepare workflow?: } 10 minutes
  \item {\bf How much time is needed to complete experiments (approximately)?: } 30 minutes
  \item {\bf Publicly available?: } \url{https://hub.docker.com/repository/docker/yipenghuang0302/quantum\_knowledge\_compilation} 
  \item {\bf Archived (provide DOI)?: } \begin{verbatim}https://doi.org/10.5281/zenodo.4321945\end{verbatim}
\end{itemize}
}

\subsection{Description}

\subsubsection{How to Access}

Our experiment requires setting up Docker (\url{https://docs.docker.com/get-started/}).
The Docker container requires about 4 GB of free disk space.
We provide three ways to access our experiment environment:
\begin{enumerate}
    \item Pulling the latest Docker image from Docker Hub (recommended);
    \item Downloading Docker image tarball from Zenodo (for artifact archiving purposes);
    \item Building a new image from a Dockerfile which pulls from GitHub repositories (demonstrates how to replicate the experiment environment).
\end{enumerate} 
Below, we provide instructions for accessing our artifact via each approach.

\paragraph{Pulling from Docker Hub}
From the Unix command line:
\begin{lstlisting}[language=bash,caption={}]
$ docker pull yipenghuang0302/\
quantum_knowledge_compilation:latest
\end{lstlisting}

\paragraph{Downloading from Zenodo archive}
\begin{enumerate}
    \item Obtain the Docker image tarball from Zenodo at this DOI:  \begin{verbatim}https://doi.org/10.5281/zenodo.4321945\end{verbatim}
    \item Then, load the Docker image tarball:
\begin{lstlisting}[language=bash,caption={}]
$ docker load --input \
quantum_knowledge_compilation.tar.gz
\end{lstlisting}

\end{enumerate}

\paragraph{Building from Dockerfile}
\begin{enumerate}
    \item Download the Ace compiler from \url{http://reasoning.cs.ucla.edu/ace/download.php}.
    \item Obtain the Dockerfile on GitHub:
\begin{lstlisting}[language=bash,caption={}]
$ git clone \
https://github.com/ \
yipenghuang0302/Cirq.git
\end{lstlisting}
    \item Place the Ace compiler tarball in the same directory as the Dockerfile:
\begin{lstlisting}[language=bash,caption={}]
$ mv ace_v3.0_linux86.tar.gz \
Cirq/kc_examples
\end{lstlisting}
    \item Change your working directory to the same directory as the Dockerfile, and build the Docker image:
\begin{lstlisting}[language=bash,caption={}]
$ cd Cirq/kc_examples
$ docker build \
--tag quantum_knowledge_compilation .
\end{lstlisting}
\end{enumerate}

\subsubsection{Hardware Dependencies}
\begin{itemize}
    \item We tested our experiment artifact on a Linux desktop and on an Apple MacBook Pro laptop (2019, with 16 GB of RAM).
    \item The full experimental results involving the largest problem instances in our paper were done on a Linux server (Two Intel Skylake Xeon Gold 6148 CPUs @ 2.40 GHz and 1 TB of RAM).
\end{itemize}

\subsubsection{Software Dependencies}
As you can see from the Dockerfile, the experiment relies on various pieces of software outlined in our checklist:
\begin{itemize}
    \item Google Cirq, a quantum programming framework: \url{https://github.com/quantumlib/Cirq}
    \item bayes-to-cnf, a Bayesian network compiler that outputs conjunctive normal form logical formulas: \url{https://github.com/gisodal/bayes-to-cnf}
    \item UCLA Ace compiler, a knowledge compilation tool that supports efficient repeated inference: \url{http://reasoning.cs.ucla.edu/ace/}
    \item Google qsim, a quantum circuit simulator based on matrix vector multiplication, which we compare against as a baseline: \url{https://github.com/quantumlib/qsim}
    \item qTorch, a quantum circuit simulator based on tensor network contraction, which we compare against as a baseline: \url{https://github.com/aspuru-guzik-group/qtorch}
\end{itemize}
These software dependencies are automatically downloaded and compiled in the Docker image.

\subsubsection{Data Sets}
We use a benchmark suite of quantum circuits in Google Cirq to validate our simulator and to measure its performance.
These input data sets are included in the Google Cirq repository, and are automatically downloaded in the Docker image.


\subsection{Installation}
No special installation is needed after you have obtained the Docker image via one of the sources above. Just enter the Docker container:
\begin{lstlisting}[language=bash,caption={}]
$ docker run -ti \
-v $(pwd):/common/home/yh804/research/pdfs \
yipenghuang0302/\
quantum_knowledge_compilation:latest
\end{lstlisting}
\begin{itemize}
    \item The \texttt{-ti} flag makes the Docker container interactive.
    \item The \texttt{-v} flag binds the directory\newline \texttt{/common/home/yh804/research/pdfs/} inside the Docker container to the present working directory of the host machine. We will be moving the experiment output plots to the host machine through this volume binding.
\end{itemize}

\subsection{Experiment Workflow}

We will perform two types of experiments in this artifact demonstration.
First, we will run a set of tests that validate the correctness of our proposed quantum circuit simulator.
Second, we will evaluate the performance of our simulation approach against three baseline quantum circuit simulators from prior work.

\subsubsection{Validation}
From the directory that you first arrive in the Docker container (\texttt{/common/home/yh804/research/}), run the following Python test suites, which should take less than 5 minutes:

\begin{lstlisting}[language=bash,caption={}]
$ pytest Google/Cirq/ \
cirq/sim/kc_sparse_simulator_test.py
$ pytest Google/Cirq/ \
kc_examples/kc_examples_test.py
\end{lstlisting}

\subsubsection{Performance Evaluation}
From the directory that you first arrive in the Docker container (\texttt{/common/home/yh804/research/}), run the following performance benchmarking experiments, which should take less than 20 minutes combined.

\begin{lstlisting}[language=bash,caption={}]
$ python3 Google/Cirq/kc_examples/\
kc_qtorch_qaoa/kc_qtorch_qaoa.py 
$ python3 Google/Cirq/kc_examples/\
kc_qtorch_vqe/kc_qtorch_vqe.py 
$ python3 Google/Cirq/kc_examples/\
kc_noise_qaoa/kc_noise_qaoa.py 
$ python3 Google/Cirq/kc_examples/\
kc_noise_vqe/kc_noise_vqe.py 
\end{lstlisting}

After running the performance benchmarks, move the experimental result PDF files to the host machine through the volume binding:

\begin{lstlisting}[language=bash,caption={}]
$ mv *.pdf /common/home/yh804/research/pdfs/
\end{lstlisting}

\subsection{Evaluation and Expected Result}

Here we describe the validation test suites and their expected results.
We then describe the expected trends in the performance benchmarking results.

\subsubsection{Validation}
We ran two Python test suites during the validation step, and both should return no errors. They test the following:
\paragraph{kc\_sparse\_simulator\_test} We show that our simulation approach passes test cases that the Google Cirq framework uses to validate simulator back ends.
\paragraph{kc\_examples\_test} We show that our simulation approach gives correct results for a benchmark suite of quantum algorithms implemented in Google Cirq, including:
\begin{enumerate}
    \item Bernstein-Vazirani algorithm
    \item Bell state creation
    \item Bell inequality
    \item Deutsch's algorithm
    \item Grover's algorithm
    \item Hidden shift algorithm
    \item Simon's algorithm
    \item Quantum Fourier transform
    \item Quantum teleportation
\end{enumerate}

\subsubsection{Performance Evaluation}
\begin{figure}[h]
\centering
\includegraphics[width=\linewidth]{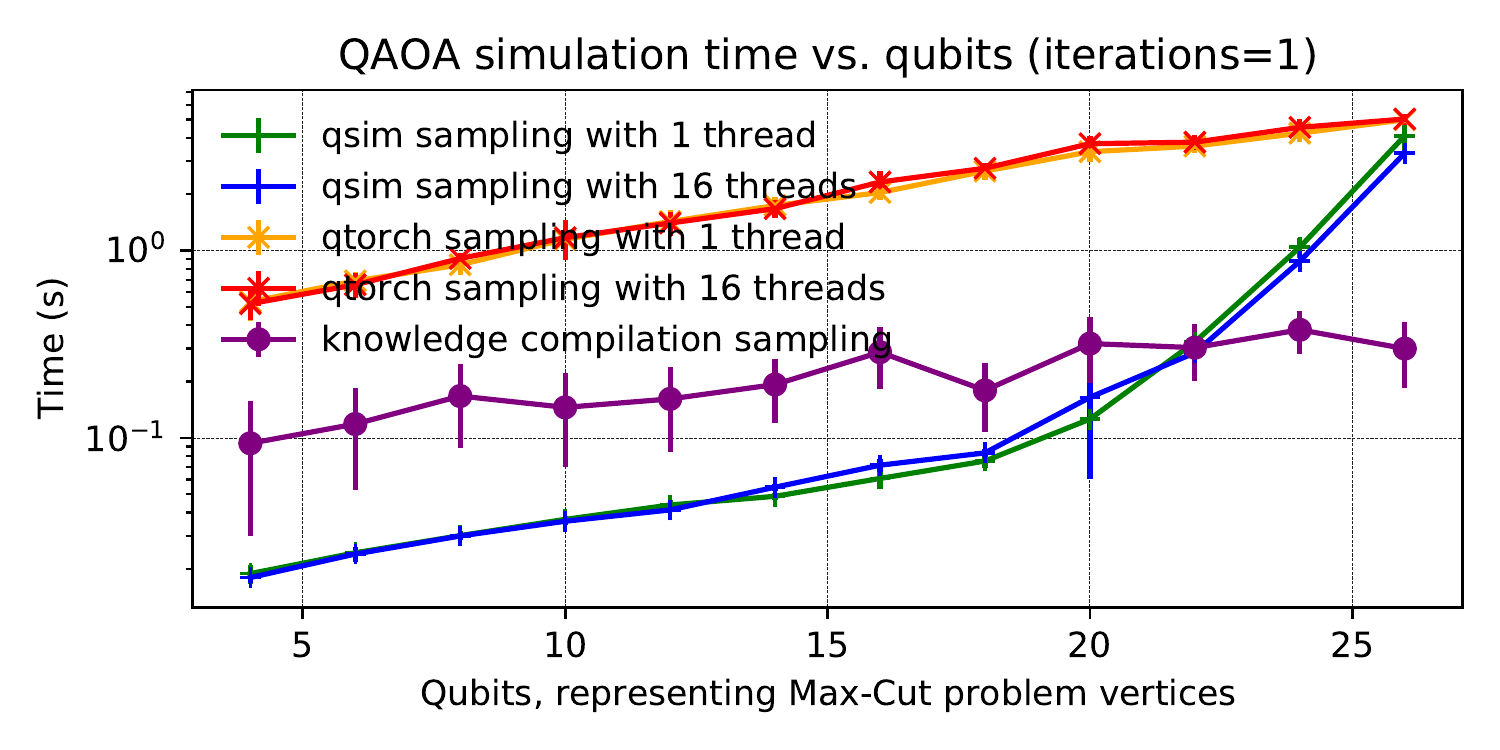}
\caption{Example output of \texttt{kc\_qtorch\_qaoa.py}: time to sample outputs from noise-free circuits for the QAOA quantum algorithm.}
\label{fig:kc_qtorch_qaoa}
\end{figure}

The performance plot PDF files replicate the results presented in Figures~\ref{fig:ideal} (sampling from noise-free quantum circuits) and~\ref{fig:noisy} (sampling from noisy quantum circuits) of the paper submission, albeit at reduced input sizes to reduce time and hardware requirements. The problem sizes are large enough to show that our simulation approach has an advantage vs. the baseline simulators. An example result is included here in Figure~\ref{fig:kc_qtorch_qaoa}.

\subsection{Experiment Customization}
The experiment input size parameters can be adjusted in the Python files \texttt{kc\_qtorch\_qaoa.py}, \texttt{kc\_qtorch\_vqe.py}, \texttt{kc\_noise\_qaoa.py}, and \texttt{kc\_noise\_vqe.py}:
\begin{itemize}
    \item The \texttt{max\_length} parameter controls the quantum circuit width, the number of qubits in the input circuit.
    \item The \texttt{p} or the \texttt{step} parameter controls the quantum circuit depth, the number iterations of the input circuit.
\end{itemize}


\subsection{Methodology}

Submission, reviewing and badging methodology:

\begin{itemize}
  \item \href{https://www.acm.org/publications/policies/artifact-review-badging}{www.acm.org/publications/policies/artifact-review-badging}
  \item \href{https://ctuning.org/ae/submission-20201122.html}{cTuning.org/ae/submission-20201122.html}
  \item \href{https://ctuning.org/ae/reviewing-20201122.html}{cTuning.org/ae/reviewing-20201122.html}
\end{itemize}


\bibliographystyle{ACM-Reference-Format}
\bibliography{main}


\begin{thebibliography}{66}


\ifx \showCODEN    \undefined \def \showCODEN     #1{\unskip}     \fi
\ifx \showDOI      \undefined \def \showDOI       #1{#1}\fi
\ifx \showISBNx    \undefined \def \showISBNx     #1{\unskip}     \fi
\ifx \showISBNxiii \undefined \def \showISBNxiii  #1{\unskip}     \fi
\ifx \showISSN     \undefined \def \showISSN      #1{\unskip}     \fi
\ifx \showLCCN     \undefined \def \showLCCN      #1{\unskip}     \fi
\ifx \shownote     \undefined \def \shownote      #1{#1}          \fi
\ifx \showarticletitle \undefined \def \showarticletitle #1{#1}   \fi
\ifx \showURL      \undefined \def \showURL       {\relax}        \fi
\providecommand\bibfield[2]{#2}
\providecommand\bibinfo[2]{#2}
\providecommand\natexlab[1]{#1}
\providecommand\showeprint[2][]{arXiv:#2}

\bibitem[\protect\citeauthoryear{{Allen}, {Barrett}, {Horsman}, {Lee}, and
  {Spekkens}}{{Allen} et~al\mbox{.}}{2017}]%
        {2017PhRvX...7c1021A}
\bibfield{author}{\bibinfo{person}{John-Mark~A. {Allen}},
  \bibinfo{person}{Jonathan {Barrett}}, \bibinfo{person}{Dominic~C. {Horsman}},
  \bibinfo{person}{Ciar{\'a}n~M. {Lee}}, {and} \bibinfo{person}{Robert~W.
  {Spekkens}}.} \bibinfo{year}{2017}\natexlab{}.
\newblock \showarticletitle{{Quantum Common Causes and Quantum Causal Models}}.
\newblock \bibinfo{journal}{\emph{Physical Review X}} \bibinfo{volume}{7},
  \bibinfo{number}{3}, Article \bibinfo{articleno}{031021} (\bibinfo{date}{Jul}
  \bibinfo{year}{2017}), \bibinfo{numpages}{031021}~pages.
\newblock
\urldef\tempurl%
\url{https://doi.org/10.1103/PhysRevX.7.031021}
\showDOI{\tempurl}
\showeprint[arxiv]{1609.09487}~[quant-ph]


\bibitem[\protect\citeauthoryear{Anschuetz, Olson, Aspuru-Guzik, and
  Cao}{Anschuetz et~al\mbox{.}}{2019}]%
        {10.1007/978-3-030-14082-3_7}
\bibfield{author}{\bibinfo{person}{Eric Anschuetz}, \bibinfo{person}{Jonathan
  Olson}, \bibinfo{person}{Al{\'a}n Aspuru-Guzik}, {and}
  \bibinfo{person}{Yudong Cao}.} \bibinfo{year}{2019}\natexlab{}.
\newblock \showarticletitle{Variational Quantum Factoring}. In
  \bibinfo{booktitle}{\emph{Quantum Technology and Optimization Problems}},
  \bibfield{editor}{\bibinfo{person}{Sebastian Feld} {and}
  \bibinfo{person}{Claudia Linnhoff-Popien}} (Eds.).
  \bibinfo{publisher}{Springer International Publishing},
  \bibinfo{address}{Cham}, \bibinfo{pages}{74--85}.
\newblock
\showISBNx{978-3-030-14082-3}


\bibitem[\protect\citeauthoryear{Arute, Arya, Babbush, Bacon, Bardin, Barends,
  Biswas, Boixo, Brandao, Buell, Burkett, Chen, Chen, Chiaro, Collins,
  Courtney, Dunsworth, Farhi, Foxen, Fowler, Gidney, Giustina, Graff, Guerin,
  Habegger, Harrigan, Hartmann, Ho, Hoffmann, Huang, Humble, Isakov, Jeffrey,
  Jiang, Kafri, Kechedzhi, Kelly, Klimov, Knysh, Korotkov, Kostritsa, Landhuis,
  Lindmark, Lucero, Lyakh, Mandr{\`a}, McClean, McEwen, Megrant, Mi,
  Michielsen, Mohseni, Mutus, Naaman, Neeley, Neill, Niu, Ostby, Petukhov,
  Platt, Quintana, Rieffel, Roushan, Rubin, Sank, Satzinger, Smelyanskiy, Sung,
  Trevithick, Vainsencher, Villalonga, White, Yao, Yeh, Zalcman, Neven, and
  Martinis}{Arute et~al\mbox{.}}{2019}]%
        {google_quantum_supremacy}
\bibfield{author}{\bibinfo{person}{Frank Arute}, \bibinfo{person}{Kunal Arya},
  \bibinfo{person}{Ryan Babbush}, \bibinfo{person}{Dave Bacon},
  \bibinfo{person}{Joseph~C. Bardin}, \bibinfo{person}{Rami Barends},
  \bibinfo{person}{Rupak Biswas}, \bibinfo{person}{Sergio Boixo},
  \bibinfo{person}{Fernando G. S.~L. Brandao}, \bibinfo{person}{David~A.
  Buell}, \bibinfo{person}{Brian Burkett}, \bibinfo{person}{Yu Chen},
  \bibinfo{person}{Zijun Chen}, \bibinfo{person}{Ben Chiaro},
  \bibinfo{person}{Roberto Collins}, \bibinfo{person}{William Courtney},
  \bibinfo{person}{Andrew Dunsworth}, \bibinfo{person}{Edward Farhi},
  \bibinfo{person}{Brooks Foxen}, \bibinfo{person}{Austin Fowler},
  \bibinfo{person}{Craig Gidney}, \bibinfo{person}{Marissa Giustina},
  \bibinfo{person}{Rob Graff}, \bibinfo{person}{Keith Guerin},
  \bibinfo{person}{Steve Habegger}, \bibinfo{person}{Matthew~P. Harrigan},
  \bibinfo{person}{Michael~J. Hartmann}, \bibinfo{person}{Alan Ho},
  \bibinfo{person}{Markus Hoffmann}, \bibinfo{person}{Trent Huang},
  \bibinfo{person}{Travis~S. Humble}, \bibinfo{person}{Sergei~V. Isakov},
  \bibinfo{person}{Evan Jeffrey}, \bibinfo{person}{Zhang Jiang},
  \bibinfo{person}{Dvir Kafri}, \bibinfo{person}{Kostyantyn Kechedzhi},
  \bibinfo{person}{Julian Kelly}, \bibinfo{person}{Paul~V. Klimov},
  \bibinfo{person}{Sergey Knysh}, \bibinfo{person}{Alexander Korotkov},
  \bibinfo{person}{Fedor Kostritsa}, \bibinfo{person}{David Landhuis},
  \bibinfo{person}{Mike Lindmark}, \bibinfo{person}{Erik Lucero},
  \bibinfo{person}{Dmitry Lyakh}, \bibinfo{person}{Salvatore Mandr{\`a}},
  \bibinfo{person}{Jarrod~R. McClean}, \bibinfo{person}{Matthew McEwen},
  \bibinfo{person}{Anthony Megrant}, \bibinfo{person}{Xiao Mi},
  \bibinfo{person}{Kristel Michielsen}, \bibinfo{person}{Masoud Mohseni},
  \bibinfo{person}{Josh Mutus}, \bibinfo{person}{Ofer Naaman},
  \bibinfo{person}{Matthew Neeley}, \bibinfo{person}{Charles Neill},
  \bibinfo{person}{Murphy~Yuezhen Niu}, \bibinfo{person}{Eric Ostby},
  \bibinfo{person}{Andre Petukhov}, \bibinfo{person}{John~C. Platt},
  \bibinfo{person}{Chris Quintana}, \bibinfo{person}{Eleanor~G. Rieffel},
  \bibinfo{person}{Pedram Roushan}, \bibinfo{person}{Nicholas~C. Rubin},
  \bibinfo{person}{Daniel Sank}, \bibinfo{person}{Kevin~J. Satzinger},
  \bibinfo{person}{Vadim Smelyanskiy}, \bibinfo{person}{Kevin~J. Sung},
  \bibinfo{person}{Matthew~D. Trevithick}, \bibinfo{person}{Amit Vainsencher},
  \bibinfo{person}{Benjamin Villalonga}, \bibinfo{person}{Theodore White},
  \bibinfo{person}{Z.~Jamie Yao}, \bibinfo{person}{Ping Yeh},
  \bibinfo{person}{Adam Zalcman}, \bibinfo{person}{Hartmut Neven}, {and}
  \bibinfo{person}{John~M. Martinis}.} \bibinfo{year}{2019}\natexlab{}.
\newblock \showarticletitle{Quantum supremacy using a programmable
  superconducting processor}.
\newblock \bibinfo{journal}{\emph{Nature}} \bibinfo{volume}{574},
  \bibinfo{number}{7779} (\bibinfo{year}{2019}), \bibinfo{pages}{505--510}.
\newblock
\showISBNx{1476-4687}
\urldef\tempurl%
\url{https://doi.org/10.1038/s41586-019-1666-5}
\showDOI{\tempurl}


\bibitem[\protect\citeauthoryear{Arute, Arya, Babbush, Bacon, Bardin, Barends,
  Boixo, Broughton, Buckley, Buell, Burkett, Bushnell, Chen, Chen, Chiaro,
  Collins, Courtney, Demura, Dunsworth, Farhi, Fowler, Foxen, Gidney, Giustina,
  Graff, Habegger, Harrigan, Ho, Hong, Huang, Ioffe, Isakov, Jeffrey, Jiang,
  Jones, Kafri, Kechedzhi, Kelly, Kim, Klimov, Korotkov, Kostritsa, Landhuis,
  Laptev, Lindmark, Leib, Lucero, Martin, Martinis, McClean, McEwen, Megrant,
  Mi, Mohseni, Mruczkiewicz, Mutus, Naaman, Neeley, Neill, Neukart, Neven, Niu,
  O'Brien, O'Gorman, Ostby, Petukhov, Putterman, Quintana, Roushan, Rubin,
  Sank, Satzinger, Skolik, Smelyanskiy, Strain, Streif, Sung, Szalay,
  Vainsencher, White, Yao, Yeh, Zalcman, and Zhou}{Arute et~al\mbox{.}}{2020}]%
        {arute2020quantum}
\bibfield{author}{\bibinfo{person}{Frank Arute}, \bibinfo{person}{Kunal Arya},
  \bibinfo{person}{Ryan Babbush}, \bibinfo{person}{Dave Bacon},
  \bibinfo{person}{Joseph~C. Bardin}, \bibinfo{person}{Rami Barends},
  \bibinfo{person}{Sergio Boixo}, \bibinfo{person}{Michael Broughton},
  \bibinfo{person}{Bob~B. Buckley}, \bibinfo{person}{David~A. Buell},
  \bibinfo{person}{Brian Burkett}, \bibinfo{person}{Nicholas Bushnell},
  \bibinfo{person}{Yu Chen}, \bibinfo{person}{Zijun Chen}, \bibinfo{person}{Ben
  Chiaro}, \bibinfo{person}{Roberto Collins}, \bibinfo{person}{William
  Courtney}, \bibinfo{person}{Sean Demura}, \bibinfo{person}{Andrew Dunsworth},
  \bibinfo{person}{Edward Farhi}, \bibinfo{person}{Austin Fowler},
  \bibinfo{person}{Brooks Foxen}, \bibinfo{person}{Craig Gidney},
  \bibinfo{person}{Marissa Giustina}, \bibinfo{person}{Rob Graff},
  \bibinfo{person}{Steve Habegger}, \bibinfo{person}{Matthew~P. Harrigan},
  \bibinfo{person}{Alan Ho}, \bibinfo{person}{Sabrina Hong},
  \bibinfo{person}{Trent Huang}, \bibinfo{person}{L.~B. Ioffe},
  \bibinfo{person}{Sergei~V. Isakov}, \bibinfo{person}{Evan Jeffrey},
  \bibinfo{person}{Zhang Jiang}, \bibinfo{person}{Cody Jones},
  \bibinfo{person}{Dvir Kafri}, \bibinfo{person}{Kostyantyn Kechedzhi},
  \bibinfo{person}{Julian Kelly}, \bibinfo{person}{Seon Kim},
  \bibinfo{person}{Paul~V. Klimov}, \bibinfo{person}{Alexander~N. Korotkov},
  \bibinfo{person}{Fedor Kostritsa}, \bibinfo{person}{David Landhuis},
  \bibinfo{person}{Pavel Laptev}, \bibinfo{person}{Mike Lindmark},
  \bibinfo{person}{Martin Leib}, \bibinfo{person}{Erik Lucero},
  \bibinfo{person}{Orion Martin}, \bibinfo{person}{John~M. Martinis},
  \bibinfo{person}{Jarrod~R. McClean}, \bibinfo{person}{Matt McEwen},
  \bibinfo{person}{Anthony Megrant}, \bibinfo{person}{Xiao Mi},
  \bibinfo{person}{Masoud Mohseni}, \bibinfo{person}{Wojciech Mruczkiewicz},
  \bibinfo{person}{Josh Mutus}, \bibinfo{person}{Ofer Naaman},
  \bibinfo{person}{Matthew Neeley}, \bibinfo{person}{Charles Neill},
  \bibinfo{person}{Florian Neukart}, \bibinfo{person}{Hartmut Neven},
  \bibinfo{person}{Murphy~Yuezhen Niu}, \bibinfo{person}{Thomas~E. O'Brien},
  \bibinfo{person}{Bryan O'Gorman}, \bibinfo{person}{Eric Ostby},
  \bibinfo{person}{Andre Petukhov}, \bibinfo{person}{Harald Putterman},
  \bibinfo{person}{Chris Quintana}, \bibinfo{person}{Pedram Roushan},
  \bibinfo{person}{Nicholas~C. Rubin}, \bibinfo{person}{Daniel Sank},
  \bibinfo{person}{Kevin~J. Satzinger}, \bibinfo{person}{Andrea Skolik},
  \bibinfo{person}{Vadim Smelyanskiy}, \bibinfo{person}{Doug Strain},
  \bibinfo{person}{Michael Streif}, \bibinfo{person}{Kevin~J. Sung},
  \bibinfo{person}{Marco Szalay}, \bibinfo{person}{Amit Vainsencher},
  \bibinfo{person}{Theodore White}, \bibinfo{person}{Z.~Jamie Yao},
  \bibinfo{person}{Ping Yeh}, \bibinfo{person}{Adam Zalcman}, {and}
  \bibinfo{person}{Leo Zhou}.} \bibinfo{year}{2020}\natexlab{}.
\newblock \bibinfo{title}{Quantum Approximate Optimization of Non-Planar Graph
  Problems on a Planar Superconducting Processor}.
\newblock
\newblock
\showeprint[arxiv]{2004.04197}~[quant-ph]


\bibitem[\protect\citeauthoryear{Barahona}{Barahona}{1982}]%
        {Barahona_1982}
\bibfield{author}{\bibinfo{person}{F Barahona}.}
  \bibinfo{year}{1982}\natexlab{}.
\newblock \showarticletitle{On the computational complexity of {Ising} spin
  glass models}.
\newblock \bibinfo{journal}{\emph{Journal of Physics A: Mathematical and
  General}} \bibinfo{volume}{15}, \bibinfo{number}{10} (\bibinfo{date}{oct}
  \bibinfo{year}{1982}), \bibinfo{pages}{3241--3253}.
\newblock
\urldef\tempurl%
\url{https://doi.org/10.1088/0305-4470/15/10/028}
\showDOI{\tempurl}


\bibitem[\protect\citeauthoryear{Beauregard}{Beauregard}{2003}]%
        {beauregard}
\bibfield{author}{\bibinfo{person}{Stephane Beauregard}.}
  \bibinfo{year}{2003}\natexlab{}.
\newblock \showarticletitle{Circuit for {Shor's} Algorithm Using 2N+3 Qubits}.
\newblock \bibinfo{journal}{\emph{Quantum Info. Comput.}} \bibinfo{volume}{3},
  \bibinfo{number}{2} (\bibinfo{date}{March} \bibinfo{year}{2003}),
  \bibinfo{pages}{175--185}.
\newblock
\showISSN{1533-7146}
\urldef\tempurl%
\url{http://dl.acm.org/citation.cfm?id=2011517.2011525}
\showURL{%
\tempurl}


\bibitem[\protect\citeauthoryear{Bernstein and Vazirani}{Bernstein and
  Vazirani}{1997}]%
        {doi:10.1137/S0097539796300921}
\bibfield{author}{\bibinfo{person}{Ethan Bernstein} {and}
  \bibinfo{person}{Umesh Vazirani}.} \bibinfo{year}{1997}\natexlab{}.
\newblock \showarticletitle{Quantum Complexity Theory}.
\newblock \bibinfo{journal}{\emph{SIAM J. Comput.}} \bibinfo{volume}{26},
  \bibinfo{number}{5} (\bibinfo{year}{1997}), \bibinfo{pages}{1411--1473}.
\newblock
\urldef\tempurl%
\url{https://doi.org/10.1137/S0097539796300921}
\showDOI{\tempurl}
\showeprint{https://doi.org/10.1137/S0097539796300921}


\bibitem[\protect\citeauthoryear{{Biamonte} and {Bergholm}}{{Biamonte} and
  {Bergholm}}{2017}]%
        {2017arXiv170800006B}
\bibfield{author}{\bibinfo{person}{Jacob {Biamonte}} {and}
  \bibinfo{person}{Ville {Bergholm}}.} \bibinfo{year}{2017}\natexlab{}.
\newblock \showarticletitle{{Tensor Networks in a Nutshell}}.
\newblock \bibinfo{journal}{\emph{arXiv e-prints}}, Article
  \bibinfo{articleno}{arXiv:1708.00006} (\bibinfo{date}{Jul}
  \bibinfo{year}{2017}), \bibinfo{numpages}{arXiv:1708.00006}~pages.
\newblock
\showeprint[arxiv]{1708.00006}~[quant-ph]


\bibitem[\protect\citeauthoryear{Boixo, Isakov, Smelyanskiy, Babbush, Ding,
  Jiang, Bremner, Martinis, and Neven}{Boixo et~al\mbox{.}}{2018}]%
        {boixo2018supremacy}
\bibfield{author}{\bibinfo{person}{Sergio Boixo}, \bibinfo{person}{Sergei~V.
  Isakov}, \bibinfo{person}{Vadim~N. Smelyanskiy}, \bibinfo{person}{Ryan
  Babbush}, \bibinfo{person}{Nan Ding}, \bibinfo{person}{Zhang Jiang},
  \bibinfo{person}{Michael~J. Bremner}, \bibinfo{person}{John~M. Martinis},
  {and} \bibinfo{person}{Hartmut Neven}.} \bibinfo{year}{2018}\natexlab{}.
\newblock \showarticletitle{Characterizing quantum supremacy in near-term
  devices}.
\newblock \bibinfo{journal}{\emph{Nature Physics}} \bibinfo{volume}{14},
  \bibinfo{number}{6} (\bibinfo{year}{2018}), \bibinfo{pages}{595--600}.
\newblock
\showISBNx{1745-2481}
\urldef\tempurl%
\url{https://doi.org/10.1038/s41567-018-0124-x}
\showDOI{\tempurl}


\bibitem[\protect\citeauthoryear{Boixo, Isakov, Smelyanskiy, and Neven}{Boixo
  et~al\mbox{.}}{2017}]%
        {boixo2017simulation}
\bibfield{author}{\bibinfo{person}{Sergio Boixo}, \bibinfo{person}{Sergei~V
  Isakov}, \bibinfo{person}{Vadim~N Smelyanskiy}, {and}
  \bibinfo{person}{Hartmut Neven}.} \bibinfo{year}{2017}\natexlab{}.
\newblock \showarticletitle{Simulation of low-depth quantum circuits as complex
  undirected graphical models}.
\newblock \bibinfo{journal}{\emph{arXiv preprint arXiv:1712.05384}}
  (\bibinfo{year}{2017}).
\newblock


\bibitem[\protect\citeauthoryear{Bravo-Prieto, LaRose, Cerezo, Subasi, Cincio,
  and Coles}{Bravo-Prieto et~al\mbox{.}}{2019}]%
        {bravoprieto2019variational}
\bibfield{author}{\bibinfo{person}{Carlos Bravo-Prieto}, \bibinfo{person}{Ryan
  LaRose}, \bibinfo{person}{M. Cerezo}, \bibinfo{person}{Yigit Subasi},
  \bibinfo{person}{Lukasz Cincio}, {and} \bibinfo{person}{Patrick~J. Coles}.}
  \bibinfo{year}{2019}\natexlab{}.
\newblock \bibinfo{title}{Variational Quantum Linear Solver: A Hybrid Algorithm
  for Linear Systems}.
\newblock
\newblock
\showeprint[arxiv]{1909.05820}~[quant-ph]


\bibitem[\protect\citeauthoryear{Chavira, Allen, and Darwiche}{Chavira
  et~al\mbox{.}}{2005}]%
        {Chavira:2005:EEP:3020336.3020350}
\bibfield{author}{\bibinfo{person}{Mark Chavira}, \bibinfo{person}{David
  Allen}, {and} \bibinfo{person}{Adnan Darwiche}.}
  \bibinfo{year}{2005}\natexlab{}.
\newblock \showarticletitle{Exploiting Evidence in Probabilistic Inference}. In
  \bibinfo{booktitle}{\emph{Proceedings of the Twenty-First Conference on
  Uncertainty in Artificial Intelligence}} (Edinburgh, Scotland)
  \emph{(\bibinfo{series}{UAI'05})}. \bibinfo{publisher}{AUAI Press},
  \bibinfo{address}{Arlington, Virginia, United States},
  \bibinfo{pages}{112--119}.
\newblock
\showISBNx{0-9749039-1-4}
\urldef\tempurl%
\url{http://dl.acm.org/citation.cfm?id=3020336.3020350}
\showURL{%
\tempurl}


\bibitem[\protect\citeauthoryear{Chavira and Darwiche}{Chavira and
  Darwiche}{2005}]%
        {cd05}
\bibfield{author}{\bibinfo{person}{Mark Chavira} {and} \bibinfo{person}{Adnan
  Darwiche}.} \bibinfo{year}{2005}\natexlab{}.
\newblock \showarticletitle{Compiling {Bayesian} Networks with Local
  Structure}. In \bibinfo{booktitle}{\emph{Proceedings of the 19th
  International Joint Conference on Artificial Intelligence}} (Edinburgh,
  Scotland) \emph{(\bibinfo{series}{IJCAI'05})}. \bibinfo{publisher}{Morgan
  Kaufmann Publishers Inc.}, \bibinfo{address}{San Francisco, CA, USA},
  \bibinfo{pages}{1306--1312}.
\newblock
\urldef\tempurl%
\url{http://dl.acm.org/citation.cfm?id=1642293.1642501}
\showURL{%
\tempurl}


\bibitem[\protect\citeauthoryear{Chavira and Darwiche}{Chavira and
  Darwiche}{2006}]%
        {Chavira:2006:ECE:2165381.2165393}
\bibfield{author}{\bibinfo{person}{Mark Chavira} {and} \bibinfo{person}{Adnan
  Darwiche}.} \bibinfo{year}{2006}\natexlab{}.
\newblock \showarticletitle{Encoding {CNFs} to Empower Component Analysis}. In
  \bibinfo{booktitle}{\emph{Theory and Applications of Satisfiability Testing -
  SAT 2006}}, \bibfield{editor}{\bibinfo{person}{Armin Biere} {and}
  \bibinfo{person}{Carla~P. Gomes}} (Eds.). \bibinfo{publisher}{Springer Berlin
  Heidelberg}, \bibinfo{address}{Berlin, Heidelberg}, \bibinfo{pages}{61--74}.
\newblock
\showISBNx{978-3-540-37207-3}


\bibitem[\protect\citeauthoryear{Chavira and Darwiche}{Chavira and
  Darwiche}{2008}]%
        {Chavira:2008:PIW:1342435.1342783}
\bibfield{author}{\bibinfo{person}{Mark Chavira} {and} \bibinfo{person}{Adnan
  Darwiche}.} \bibinfo{year}{2008}\natexlab{}.
\newblock \showarticletitle{On Probabilistic Inference by Weighted Model
  Counting}.
\newblock \bibinfo{journal}{\emph{Artif. Intell.}} \bibinfo{volume}{172},
  \bibinfo{number}{6-7} (\bibinfo{date}{April} \bibinfo{year}{2008}),
  \bibinfo{pages}{772--799}.
\newblock
\showISSN{0004-3702}
\urldef\tempurl%
\url{https://doi.org/10.1016/j.artint.2007.11.002}
\showDOI{\tempurl}


\bibitem[\protect\citeauthoryear{Chib and Greenberg}{Chib and
  Greenberg}{1995}]%
        {chib1995understanding}
\bibfield{author}{\bibinfo{person}{Siddhartha Chib} {and}
  \bibinfo{person}{Edward Greenberg}.} \bibinfo{year}{1995}\natexlab{}.
\newblock \showarticletitle{Understanding the {Metropolis-Hastings} algorithm}.
\newblock \bibinfo{journal}{\emph{The american statistician}}
  \bibinfo{volume}{49}, \bibinfo{number}{4} (\bibinfo{year}{1995}),
  \bibinfo{pages}{327--335}.
\newblock


\bibitem[\protect\citeauthoryear{Choi, Vergari, and Van~den Broeck}{Choi
  et~al\mbox{.}}{2020}]%
        {ProbCirc20}
\bibfield{author}{\bibinfo{person}{YooJung Choi}, \bibinfo{person}{Antonio
  Vergari}, {and} \bibinfo{person}{Guy Van~den Broeck}.}
  \bibinfo{year}{2020}\natexlab{}.
\newblock \showarticletitle{Probabilistic Circuits: A Unifying Framework for
  Tractable Probabilistic Models}.
\newblock  (\bibinfo{year}{2020}).
\newblock


\bibitem[\protect\citeauthoryear{Clauser, Horne, Shimony, and Holt}{Clauser
  et~al\mbox{.}}{1969}]%
        {PhysRevLett.23.880}
\bibfield{author}{\bibinfo{person}{John~F. Clauser},
  \bibinfo{person}{Michael~A. Horne}, \bibinfo{person}{Abner Shimony}, {and}
  \bibinfo{person}{Richard~A. Holt}.} \bibinfo{year}{1969}\natexlab{}.
\newblock \showarticletitle{Proposed Experiment to Test Local Hidden-Variable
  Theories}.
\newblock \bibinfo{journal}{\emph{Phys. Rev. Lett.}}  \bibinfo{volume}{23}
  (\bibinfo{date}{Oct} \bibinfo{year}{1969}), \bibinfo{pages}{880--884}.
\newblock
Issue 15.
\urldef\tempurl%
\url{https://doi.org/10.1103/PhysRevLett.23.880}
\showDOI{\tempurl}


\bibitem[\protect\citeauthoryear{Cross, Bishop, Sheldon, Nation, and
  Gambetta}{Cross et~al\mbox{.}}{2019}]%
        {PhysRevA.100.032328}
\bibfield{author}{\bibinfo{person}{Andrew~W. Cross}, \bibinfo{person}{Lev~S.
  Bishop}, \bibinfo{person}{Sarah Sheldon}, \bibinfo{person}{Paul~D. Nation},
  {and} \bibinfo{person}{Jay~M. Gambetta}.} \bibinfo{year}{2019}\natexlab{}.
\newblock \showarticletitle{Validating quantum computers using randomized model
  circuits}.
\newblock \bibinfo{journal}{\emph{Phys. Rev. A}}  \bibinfo{volume}{100}
  (\bibinfo{date}{Sep} \bibinfo{year}{2019}), \bibinfo{pages}{032328}.
\newblock
Issue 3.
\urldef\tempurl%
\url{https://doi.org/10.1103/PhysRevA.100.032328}
\showDOI{\tempurl}


\bibitem[\protect\citeauthoryear{Darwiche}{Darwiche}{2002}]%
        {Darwiche:2002:LAF:3087093.3087128}
\bibfield{author}{\bibinfo{person}{Adnan Darwiche}.}
  \bibinfo{year}{2002}\natexlab{}.
\newblock \showarticletitle{A Logical Approach to Factoring Belief Networks}.
  In \bibinfo{booktitle}{\emph{Proceedings of the Eights International
  Conference on Principles of Knowledge Representation and Reasoning}}
  (Toulouse, France) \emph{(\bibinfo{series}{KR'02})}.
  \bibinfo{publisher}{Morgan Kaufmann Publishers Inc.}, \bibinfo{address}{San
  Francisco, CA, USA}, \bibinfo{pages}{409--420}.
\newblock
\showISBNx{1-55860-554-1}
\urldef\tempurl%
\url{http://dl.acm.org/citation.cfm?id=3087093.3087128}
\showURL{%
\tempurl}


\bibitem[\protect\citeauthoryear{Darwiche}{Darwiche}{2003}]%
        {darwiche_2003}
\bibfield{author}{\bibinfo{person}{Adnan Darwiche}.}
  \bibinfo{year}{2003}\natexlab{}.
\newblock \showarticletitle{A Differential Approach to Inference in {Bayesian}
  Networks}.
\newblock \bibinfo{journal}{\emph{J. ACM}} \bibinfo{volume}{50},
  \bibinfo{number}{3} (\bibinfo{date}{May} \bibinfo{year}{2003}),
  \bibinfo{pages}{280--305}.
\newblock
\showISSN{0004-5411}
\urldef\tempurl%
\url{https://doi.org/10.1145/765568.765570}
\showDOI{\tempurl}


\bibitem[\protect\citeauthoryear{Darwiche}{Darwiche}{2009}]%
        {darwiche2009modeling}
\bibfield{author}{\bibinfo{person}{Adnan Darwiche}.}
  \bibinfo{year}{2009}\natexlab{}.
\newblock \bibinfo{booktitle}{\emph{Modeling and Reasoning with Bayesian
  Networks} (\bibinfo{edition}{1st} ed.)}.
\newblock \bibinfo{publisher}{Cambridge University Press},
  \bibinfo{address}{New York, NY, USA}.
\newblock
\showISBNx{0521884381, 9780521884389}


\bibitem[\protect\citeauthoryear{Darwiche and Marquis}{Darwiche and
  Marquis}{2002}]%
        {darwiche2002knowledge}
\bibfield{author}{\bibinfo{person}{Adnan Darwiche} {and}
  \bibinfo{person}{Pierre Marquis}.} \bibinfo{year}{2002}\natexlab{}.
\newblock \showarticletitle{A knowledge compilation map}.
\newblock \bibinfo{journal}{\emph{Journal of Artificial Intelligence Research}}
   \bibinfo{volume}{17} (\bibinfo{year}{2002}), \bibinfo{pages}{229--264}.
\newblock


\bibitem[\protect\citeauthoryear{Deutsch}{Deutsch}{1989}]%
        {Deutsch_Networks}
\bibfield{author}{\bibinfo{person}{D. Deutsch}.}
  \bibinfo{year}{1989}\natexlab{}.
\newblock \showarticletitle{Quantum Computational Networks}.
\newblock \bibinfo{journal}{\emph{Proceedings of the Royal Society of London.
  Series A, Mathematical and Physical Sciences}} \bibinfo{volume}{425},
  \bibinfo{number}{1868} (\bibinfo{year}{1989}), \bibinfo{pages}{73--90}.
\newblock
\showISSN{00804630}
\urldef\tempurl%
\url{http://www.jstor.org/stable/2398494}
\showURL{%
\tempurl}


\bibitem[\protect\citeauthoryear{{Deutsch} and {Jozsa}}{{Deutsch} and
  {Jozsa}}{1992}]%
        {1992RSPSA.439..553D}
\bibfield{author}{\bibinfo{person}{David {Deutsch}} {and}
  \bibinfo{person}{Richard {Jozsa}}.} \bibinfo{year}{1992}\natexlab{}.
\newblock \showarticletitle{{Rapid Solution of Problems by Quantum
  Computation}}.
\newblock \bibinfo{journal}{\emph{Proceedings of the Royal Society of London
  Series A}} \bibinfo{volume}{439}, \bibinfo{number}{1907}
  (\bibinfo{date}{Dec.} \bibinfo{year}{1992}), \bibinfo{pages}{553--558}.
\newblock
\urldef\tempurl%
\url{https://doi.org/10.1098/rspa.1992.0167}
\showDOI{\tempurl}


\bibitem[\protect\citeauthoryear{{Farhi}, {Goldstone}, and {Gutmann}}{{Farhi}
  et~al\mbox{.}}{2014}]%
        {farhi_qaoa}
\bibfield{author}{\bibinfo{person}{Edward {Farhi}}, \bibinfo{person}{Jeffrey
  {Goldstone}}, {and} \bibinfo{person}{Sam {Gutmann}}.}
  \bibinfo{year}{2014}\natexlab{}.
\newblock \showarticletitle{{A Quantum Approximate Optimization Algorithm}}.
\newblock \bibinfo{journal}{\emph{arXiv e-prints}}, Article
  \bibinfo{articleno}{arXiv:1411.4028} (\bibinfo{date}{Nov}
  \bibinfo{year}{2014}), \bibinfo{numpages}{arXiv:1411.4028}~pages.
\newblock
\showeprint[arxiv]{1411.4028}~[quant-ph]


\bibitem[\protect\citeauthoryear{Farhi and Harrow}{Farhi and Harrow}{2016}]%
        {farhi2016quantum}
\bibfield{author}{\bibinfo{person}{Edward Farhi} {and} \bibinfo{person}{Aram~W
  Harrow}.} \bibinfo{year}{2016}\natexlab{}.
\newblock \bibinfo{title}{Quantum Supremacy through the Quantum Approximate
  Optimization Algorithm}.
\newblock
\newblock
\showeprint[arxiv]{1602.07674}~[quant-ph]


\bibitem[\protect\citeauthoryear{Feynman}{Feynman}{2006}]%
        {feynman2006qed}
\bibfield{author}{\bibinfo{person}{Richard~Phillips Feynman}.}
  \bibinfo{year}{2006}\natexlab{}.
\newblock \bibinfo{booktitle}{\emph{{QED}: {The} strange theory of light and
  matter}}.
\newblock \bibinfo{publisher}{Princeton University Press}.
\newblock


\bibitem[\protect\citeauthoryear{Fried, Sawaya, Cao, Kivlichan, Romero,
  Aspuru-Guzik, and Hen}{Fried et~al\mbox{.}}{2018}]%
        {2017arXiv170903636S}
\bibfield{author}{\bibinfo{person}{E.~Schuyler Fried}, \bibinfo{person}{Nicolas
  P.~D. Sawaya}, \bibinfo{person}{Yudong Cao}, \bibinfo{person}{Ian~D.
  Kivlichan}, \bibinfo{person}{Jhonathan Romero}, \bibinfo{person}{Alán
  Aspuru-Guzik}, {and} \bibinfo{person}{Itay Hen, ed.}}
  \bibinfo{year}{2018}\natexlab{}.
\newblock \showarticletitle{{qTorch}: The quantum tensor contraction handler}.
\newblock \bibinfo{journal}{\emph{PLoS ONE}} \bibinfo{volume}{13},
  \bibinfo{number}{12} (\bibinfo{date}{12} \bibinfo{year}{2018}).
\newblock
\urldef\tempurl%
\url{https://doi.org/10.1371/journal.pone.0208510}
\showDOI{\tempurl}


\bibitem[\protect\citeauthoryear{Grover}{Grover}{1996}]%
        {Grover:1996:FQM:237814.237866}
\bibfield{author}{\bibinfo{person}{Lov~K. Grover}.}
  \bibinfo{year}{1996}\natexlab{}.
\newblock \showarticletitle{A Fast Quantum Mechanical Algorithm for Database
  Search}. In \bibinfo{booktitle}{\emph{Proceedings of the Twenty-eighth Annual
  ACM Symposium on Theory of Computing}} (Philadelphia, Pennsylvania, USA)
  \emph{(\bibinfo{series}{STOC '96})}. \bibinfo{publisher}{ACM},
  \bibinfo{address}{New York, NY, USA}, \bibinfo{pages}{212--219}.
\newblock
\showISBNx{0-89791-785-5}
\urldef\tempurl%
\url{https://doi.org/10.1145/237814.237866}
\showDOI{\tempurl}


\bibitem[\protect\citeauthoryear{Grover}{Grover}{2001}]%
        {doi:10.1119/1.1359518}
\bibfield{author}{\bibinfo{person}{Lov~K. Grover}.}
  \bibinfo{year}{2001}\natexlab{}.
\newblock \showarticletitle{From {Schr\"{o}dinger's} equation to the quantum
  search algorithm}.
\newblock \bibinfo{journal}{\emph{American Journal of Physics}}
  \bibinfo{volume}{69}, \bibinfo{number}{7} (\bibinfo{year}{2001}),
  \bibinfo{pages}{769--777}.
\newblock
\urldef\tempurl%
\url{https://doi.org/10.1119/1.1359518}
\showDOI{\tempurl}
\showeprint{https://doi.org/10.1119/1.1359518}


\bibitem[\protect\citeauthoryear{Harrow and Montanaro}{Harrow and
  Montanaro}{2017}]%
        {Harrow}
\bibfield{author}{\bibinfo{person}{Aram~W. Harrow} {and}
  \bibinfo{person}{Ashley Montanaro}.} \bibinfo{year}{2017}\natexlab{}.
\newblock \showarticletitle{Quantum computational supremacy}.
\newblock \bibinfo{journal}{\emph{Nature}}  \bibinfo{volume}{549}
  (\bibinfo{date}{13 09} \bibinfo{year}{2017}), \bibinfo{pages}{203 EP --}.
\newblock
\urldef\tempurl%
\url{https://doi.org/10.1038/nature23458}
\showURL{%
\tempurl}


\bibitem[\protect\citeauthoryear{Henson, Lal, and Pusey}{Henson
  et~al\mbox{.}}{2014}]%
        {Henson_2014}
\bibfield{author}{\bibinfo{person}{Joe Henson}, \bibinfo{person}{Raymond Lal},
  {and} \bibinfo{person}{Matthew~F Pusey}.} \bibinfo{year}{2014}\natexlab{}.
\newblock \showarticletitle{Theory-independent limits on correlations from
  generalized {Bayesian} networks}.
\newblock \bibinfo{journal}{\emph{New Journal of Physics}}
  \bibinfo{volume}{16}, \bibinfo{number}{11} (\bibinfo{date}{nov}
  \bibinfo{year}{2014}), \bibinfo{pages}{113043}.
\newblock
\urldef\tempurl%
\url{https://doi.org/10.1088/1367-2630/16/11/113043}
\showDOI{\tempurl}


\bibitem[\protect\citeauthoryear{Holtzen, Millstein, and Van~den
  Broeck}{Holtzen et~al\mbox{.}}{2019}]%
        {HoltzenUAI19}
\bibfield{author}{\bibinfo{person}{Steven Holtzen}, \bibinfo{person}{Todd
  Millstein}, {and} \bibinfo{person}{Guy Van~den Broeck}.}
  \bibinfo{year}{2019}\natexlab{}.
\newblock \showarticletitle{Generating and Sampling Orbits for Lifted
  Probabilistic Inference}. In \bibinfo{booktitle}{\emph{Proceedings of the
  35th Conference on Uncertainty in Artificial Intelligence (UAI)}}.
\newblock


\bibitem[\protect\citeauthoryear{Holtzen, {Van den Broeck}, and
  Millstein}{Holtzen et~al\mbox{.}}{2020}]%
        {HoltzenOOPSLA20}
\bibfield{author}{\bibinfo{person}{Steven Holtzen}, \bibinfo{person}{Guy {Van
  den Broeck}}, {and} \bibinfo{person}{Todd Millstein}.}
  \bibinfo{year}{2020}\natexlab{}.
\newblock \showarticletitle{Scaling Exact Inference for Discrete Probabilistic
  Programs}.
\newblock \bibinfo{journal}{\emph{Proc. ACM Program. Lang. (OOPSLA)}}
  (\bibinfo{year}{2020}).
\newblock
\urldef\tempurl%
\url{https://doi.org/10.1145/342820}
\showDOI{\tempurl}


\bibitem[\protect\citeauthoryear{Huang, Bharti, and Rebentrost}{Huang
  et~al\mbox{.}}{2019}]%
        {huang2019nearterm}
\bibfield{author}{\bibinfo{person}{Hsin-Yuan Huang}, \bibinfo{person}{Kishor
  Bharti}, {and} \bibinfo{person}{Patrick Rebentrost}.}
  \bibinfo{year}{2019}\natexlab{}.
\newblock \bibinfo{title}{Near-term quantum algorithms for linear systems of
  equations}.
\newblock
\newblock
\showeprint[arxiv]{1909.07344}~[quant-ph]


\bibitem[\protect\citeauthoryear{Huang and Martonosi}{Huang and
  Martonosi}{2019a}]%
        {huang_et_al:OASIcs:2019:10196}
\bibfield{author}{\bibinfo{person}{Yipeng Huang} {and}
  \bibinfo{person}{Margaret Martonosi}.} \bibinfo{year}{2019}\natexlab{a}.
\newblock \showarticletitle{{QDB: From Quantum Algorithms Towards Correct
  Quantum Programs}}. In \bibinfo{booktitle}{\emph{9th Workshop on Evaluation
  and Usability of Programming Languages and Tools (PLATEAU 2018)}}
  \emph{(\bibinfo{series}{OpenAccess Series in Informatics (OASIcs)},
  Vol.~\bibinfo{volume}{67})}, \bibfield{editor}{\bibinfo{person}{Titus Barik},
  \bibinfo{person}{Joshua Sunshine}, {and} \bibinfo{person}{Sarah Chasins}}
  (Eds.). \bibinfo{publisher}{Schloss Dagstuhl--Leibniz-Zentrum fuer
  Informatik}, \bibinfo{address}{Dagstuhl, Germany},
  \bibinfo{pages}{4:1--4:14}.
\newblock
\showISBNx{978-3-95977-091-0}
\showISSN{2190-6807}
\urldef\tempurl%
\url{https://doi.org/10.4230/OASIcs.PLATEAU.2018.4}
\showDOI{\tempurl}


\bibitem[\protect\citeauthoryear{Huang and Martonosi}{Huang and
  Martonosi}{2019b}]%
        {huang_isca}
\bibfield{author}{\bibinfo{person}{Yipeng Huang} {and}
  \bibinfo{person}{Margaret Martonosi}.} \bibinfo{year}{2019}\natexlab{b}.
\newblock \showarticletitle{{Statistical Assertions for Validating Patterns and
  Finding Bugs in Quantum Programs}}. In \bibinfo{booktitle}{\emph{Proceedings
  of the 46th International Symposium on Computer Architecture}} (Phoenix, AZ)
  \emph{(\bibinfo{series}{ISCA '19})}.
\newblock


\bibitem[\protect\citeauthoryear{Kaye, Laflamme, and Mosca}{Kaye
  et~al\mbox{.}}{2007}]%
        {Kaye:2007:IQC:1206629}
\bibfield{author}{\bibinfo{person}{Phillip Kaye}, \bibinfo{person}{Raymond
  Laflamme}, {and} \bibinfo{person}{Michele Mosca}.}
  \bibinfo{year}{2007}\natexlab{}.
\newblock \bibinfo{booktitle}{\emph{An Introduction to Quantum Computing}}.
\newblock \bibinfo{publisher}{Oxford University Press, Inc.},
  \bibinfo{address}{New York, NY, USA}.
\newblock
\showISBNx{0198570007}


\bibitem[\protect\citeauthoryear{Kimmig, Broeck, and Raedt}{Kimmig
  et~al\mbox{.}}{2011}]%
        {10.5555/2900423.2900456}
\bibfield{author}{\bibinfo{person}{Angelika Kimmig}, \bibinfo{person}{Guy
  Van~den Broeck}, {and} \bibinfo{person}{Luc~De Raedt}.}
  \bibinfo{year}{2011}\natexlab{}.
\newblock \showarticletitle{An Algebraic Prolog for Reasoning about Possible
  Worlds}. In \bibinfo{booktitle}{\emph{Proceedings of the Twenty-Fifth AAAI
  Conference on Artificial Intelligence}} (San Francisco, California)
  \emph{(\bibinfo{series}{AAAI'11})}. \bibinfo{publisher}{AAAI Press},
  \bibinfo{pages}{209--214}.
\newblock


\bibitem[\protect\citeauthoryear{Kimmig, Van~den Broeck, and De~Raedt}{Kimmig
  et~al\mbox{.}}{2017}]%
        {Kimmig:2017:AMC:3095823.3095885}
\bibfield{author}{\bibinfo{person}{Angelika Kimmig}, \bibinfo{person}{Guy
  Van~den Broeck}, {and} \bibinfo{person}{Luc De~Raedt}.}
  \bibinfo{year}{2017}\natexlab{}.
\newblock \showarticletitle{Algebraic Model Counting}.
\newblock \bibinfo{journal}{\emph{J. of Applied Logic}} \bibinfo{volume}{22},
  \bibinfo{number}{C} (\bibinfo{date}{July} \bibinfo{year}{2017}),
  \bibinfo{pages}{46--62}.
\newblock
\showISSN{1570-8683}
\urldef\tempurl%
\url{https://doi.org/10.1016/j.jal.2016.11.031}
\showDOI{\tempurl}


\bibitem[\protect\citeauthoryear{Koller and Friedman}{Koller and
  Friedman}{2009}]%
        {Koller:2009:PGM:1795555}
\bibfield{author}{\bibinfo{person}{Daphne Koller} {and} \bibinfo{person}{Nir
  Friedman}.} \bibinfo{year}{2009}\natexlab{}.
\newblock \bibinfo{booktitle}{\emph{Probabilistic Graphical Models: Principles
  and Techniques - Adaptive Computation and Machine Learning}}.
\newblock \bibinfo{publisher}{The MIT Press}.
\newblock
\showISBNx{0262013193, 9780262013192}


\bibitem[\protect\citeauthoryear{Korb and Nicholson}{Korb and
  Nicholson}{2010}]%
        {Korb:2010:BAI:1941985}
\bibfield{author}{\bibinfo{person}{Kevin~B. Korb} {and} \bibinfo{person}{Ann~E.
  Nicholson}.} \bibinfo{year}{2010}\natexlab{}.
\newblock \bibinfo{booktitle}{\emph{Bayesian Artificial Intelligence, Second
  Edition} (\bibinfo{edition}{2nd} ed.)}.
\newblock \bibinfo{publisher}{CRC Press, Inc.}, \bibinfo{address}{Boca Raton,
  FL, USA}.
\newblock
\showISBNx{1439815917, 9781439815915}


\bibitem[\protect\citeauthoryear{{Laskey}}{{Laskey}}{1995}]%
        {384252}
\bibfield{author}{\bibinfo{person}{K.~B. {Laskey}}.}
  \bibinfo{year}{1995}\natexlab{}.
\newblock \showarticletitle{Sensitivity analysis for probability assessments in
  {Bayesian} networks}.
\newblock \bibinfo{journal}{\emph{IEEE Transactions on Systems, Man, and
  Cybernetics}} \bibinfo{volume}{25}, \bibinfo{number}{6}
  (\bibinfo{year}{1995}), \bibinfo{pages}{901--909}.
\newblock


\bibitem[\protect\citeauthoryear{Lubasch, Joo, Moinier, Kiffner, and
  Jaksch}{Lubasch et~al\mbox{.}}{2020}]%
        {PhysRevA.101.010301}
\bibfield{author}{\bibinfo{person}{Michael Lubasch}, \bibinfo{person}{Jaewoo
  Joo}, \bibinfo{person}{Pierre Moinier}, \bibinfo{person}{Martin Kiffner},
  {and} \bibinfo{person}{Dieter Jaksch}.} \bibinfo{year}{2020}\natexlab{}.
\newblock \showarticletitle{Variational quantum algorithms for nonlinear
  problems}.
\newblock \bibinfo{journal}{\emph{Phys. Rev. A}}  \bibinfo{volume}{101}
  (\bibinfo{date}{Jan} \bibinfo{year}{2020}), \bibinfo{pages}{010301}.
\newblock
Issue 1.
\urldef\tempurl%
\url{https://doi.org/10.1103/PhysRevA.101.010301}
\showDOI{\tempurl}


\bibitem[\protect\citeauthoryear{MacKay}{MacKay}{2002}]%
        {10.5555/971143}
\bibfield{author}{\bibinfo{person}{David J.~C. MacKay}.}
  \bibinfo{year}{2002}\natexlab{}.
\newblock \bibinfo{booktitle}{\emph{Information Theory, Inference \& Learning
  Algorithms}}.
\newblock \bibinfo{publisher}{Cambridge University Press},
  \bibinfo{address}{USA}.
\newblock
\showISBNx{0521642981}


\bibitem[\protect\citeauthoryear{Markov and Shi}{Markov and Shi}{2008}]%
        {doi:10.1137/050644756}
\bibfield{author}{\bibinfo{person}{I. Markov} {and} \bibinfo{person}{Y. Shi}.}
  \bibinfo{year}{2008}\natexlab{}.
\newblock \showarticletitle{Simulating Quantum Computation by Contracting
  Tensor Networks}.
\newblock \bibinfo{journal}{\emph{SIAM J. Comput.}} \bibinfo{volume}{38},
  \bibinfo{number}{3} (\bibinfo{year}{2008}), \bibinfo{pages}{963--981}.
\newblock
\urldef\tempurl%
\url{https://doi.org/10.1137/050644756}
\showDOI{\tempurl}
\showeprint{https://doi.org/10.1137/050644756}


\bibitem[\protect\citeauthoryear{Markov, Fatima, Isakov, and Boixo}{Markov
  et~al\mbox{.}}{2018}]%
        {markov2018quantum}
\bibfield{author}{\bibinfo{person}{Igor~L Markov}, \bibinfo{person}{Aneeqa
  Fatima}, \bibinfo{person}{Sergei~V Isakov}, {and} \bibinfo{person}{Sergio
  Boixo}.} \bibinfo{year}{2018}\natexlab{}.
\newblock \showarticletitle{Quantum supremacy is both closer and farther than
  it appears}.
\newblock \bibinfo{journal}{\emph{arXiv preprint arXiv:1807.10749}}
  (\bibinfo{year}{2018}).
\newblock


\bibitem[\protect\citeauthoryear{Mermin}{Mermin}{2007}]%
        {mermin2007quantum}
\bibfield{author}{\bibinfo{person}{N.D. Mermin}.}
  \bibinfo{year}{2007}\natexlab{}.
\newblock \bibinfo{booktitle}{\emph{Quantum Computer Science: An
  Introduction}}.
\newblock \bibinfo{publisher}{Cambridge University Press}.
\newblock
\showISBNx{9780521876582}
\showLCCN{2007282225}


\bibitem[\protect\citeauthoryear{{National Academies of Sciences, Engineering,
  and Medicine}}{{National Academies of Sciences, Engineering, and
  Medicine}}{2019}]%
        {NAP25196}
\bibfield{author}{\bibinfo{person}{{National Academies of Sciences,
  Engineering, and Medicine}}.} \bibinfo{year}{2019}\natexlab{}.
\newblock \bibinfo{booktitle}{\emph{Quantum Computing: Progress and
  Prospects}}.
\newblock \bibinfo{publisher}{The National Academies Press},
  \bibinfo{address}{Washington, DC}.
\newblock
\showISBNx{978-0-309-47969-1}
\urldef\tempurl%
\url{https://doi.org/10.17226/25196}
\showDOI{\tempurl}


\bibitem[\protect\citeauthoryear{Nielsen and Chuang}{Nielsen and
  Chuang}{2011}]%
        {Nielsen:2011:QCQ:1972505}
\bibfield{author}{\bibinfo{person}{Michael~A. Nielsen} {and}
  \bibinfo{person}{Isaac~L. Chuang}.} \bibinfo{year}{2011}\natexlab{}.
\newblock \bibinfo{booktitle}{\emph{Quantum Computation and Quantum
  Information: 10th Anniversary Edition} (\bibinfo{edition}{10th} ed.)}.
\newblock \bibinfo{publisher}{Cambridge University Press},
  \bibinfo{address}{New York, NY, USA}.
\newblock
\showISBNx{1107002176, 9781107002173}


\bibitem[\protect\citeauthoryear{Pearl}{Pearl}{2014}]%
        {pearl2014probabilistic}
\bibfield{author}{\bibinfo{person}{Judea Pearl}.}
  \bibinfo{year}{2014}\natexlab{}.
\newblock \bibinfo{booktitle}{\emph{Probabilistic reasoning in intelligent
  systems: networks of plausible inference}}.
\newblock \bibinfo{publisher}{Elsevier}.
\newblock


\bibitem[\protect\citeauthoryear{Peruzzo, McClean, Shadbolt, Yung, Zhou, Love,
  Aspuru-Guzik, and O'Brien}{Peruzzo et~al\mbox{.}}{2014}]%
        {vqe}
\bibfield{author}{\bibinfo{person}{Alberto Peruzzo}, \bibinfo{person}{Jarrod
  McClean}, \bibinfo{person}{Peter Shadbolt}, \bibinfo{person}{Man-Hong Yung},
  \bibinfo{person}{Xiao-Qi Zhou}, \bibinfo{person}{Peter~J. Love},
  \bibinfo{person}{Al{\'a}n Aspuru-Guzik}, {and} \bibinfo{person}{Jeremy~L.
  O'Brien}.} \bibinfo{year}{2014}\natexlab{}.
\newblock \showarticletitle{A variational eigenvalue solver on a photonic
  quantum processor}.
\newblock \bibinfo{journal}{\emph{Nature Communications}}  \bibinfo{volume}{5}
  (\bibinfo{date}{23 07} \bibinfo{year}{2014}), \bibinfo{pages}{4213 EP --}.
\newblock
\urldef\tempurl%
\url{http://dx.doi.org/10.1038/ncomms5213}
\showURL{%
\tempurl}


\bibitem[\protect\citeauthoryear{{Pienaar} and {Brukner}}{{Pienaar} and
  {Brukner}}{2015}]%
        {2015NJPh...17g3020P}
\bibfield{author}{\bibinfo{person}{Jacques {Pienaar}} {and}
  \bibinfo{person}{{\v{C}}aslav {Brukner}}.} \bibinfo{year}{2015}\natexlab{}.
\newblock \showarticletitle{{A graph-separation theorem for quantum causal
  models}}.
\newblock \bibinfo{journal}{\emph{New Journal of Physics}}
  \bibinfo{volume}{17}, \bibinfo{number}{7}, Article
  \bibinfo{articleno}{073020} (\bibinfo{date}{Jul} \bibinfo{year}{2015}),
  \bibinfo{numpages}{073020}~pages.
\newblock
\urldef\tempurl%
\url{https://doi.org/10.1088/1367-2630/17/7/073020}
\showDOI{\tempurl}
\showeprint[arxiv]{1406.0430}~[quant-ph]


\bibitem[\protect\citeauthoryear{Preskill}{Preskill}{2018}]%
        {Preskill2018quantumcomputingin}
\bibfield{author}{\bibinfo{person}{John Preskill}.}
  \bibinfo{year}{2018}\natexlab{}.
\newblock \showarticletitle{Quantum {C}omputing in the {NISQ} era and beyond}.
\newblock \bibinfo{journal}{\emph{{Quantum}}}  \bibinfo{volume}{2}
  (\bibinfo{date}{Aug.} \bibinfo{year}{2018}), \bibinfo{pages}{79}.
\newblock
\showISSN{2521-327X}
\urldef\tempurl%
\url{https://doi.org/10.22331/q-2018-08-06-79}
\showDOI{\tempurl}


\bibitem[\protect\citeauthoryear{Resch and Karpuzcu}{Resch and
  Karpuzcu}{2019}]%
        {resch2019benchmarking}
\bibfield{author}{\bibinfo{person}{Salonik Resch} {and}
  \bibinfo{person}{Ulya~R. Karpuzcu}.} \bibinfo{year}{2019}\natexlab{}.
\newblock \bibinfo{title}{Benchmarking Quantum Computers and the Impact of
  Quantum Noise}.
\newblock
\newblock
\showeprint[arxiv]{1912.00546}~[quant-ph]


\bibitem[\protect\citeauthoryear{Sang, Bearne, and Kautz}{Sang
  et~al\mbox{.}}{2005}]%
        {skb05}
\bibfield{author}{\bibinfo{person}{Tian Sang}, \bibinfo{person}{Paul Bearne},
  {and} \bibinfo{person}{Henry Kautz}.} \bibinfo{year}{2005}\natexlab{}.
\newblock \showarticletitle{Performing {Bayesian} Inference by Weighted Model
  Counting}. In \bibinfo{booktitle}{\emph{Proceedings of the 20th National
  Conference on Artificial Intelligence - Volume 1}} (Pittsburgh, Pennsylvania)
  \emph{(\bibinfo{series}{AAAI'05})}. \bibinfo{publisher}{AAAI Press},
  \bibinfo{pages}{475--481}.
\newblock
\showISBNx{1-57735-236-x}
\urldef\tempurl%
\url{http://dl.acm.org/citation.cfm?id=1619332.1619409}
\showURL{%
\tempurl}


\bibitem[\protect\citeauthoryear{Shor}{Shor}{1997}]%
        {doi:10.1137/S0097539795293172}
\bibfield{author}{\bibinfo{person}{P. Shor}.} \bibinfo{year}{1997}\natexlab{}.
\newblock \showarticletitle{Polynomial-Time Algorithms for Prime Factorization
  and Discrete Logarithms on a Quantum Computer}.
\newblock \bibinfo{journal}{\emph{SIAM J. Comput.}} \bibinfo{volume}{26},
  \bibinfo{number}{5} (\bibinfo{year}{1997}), \bibinfo{pages}{1484--1509}.
\newblock
\urldef\tempurl%
\url{https://doi.org/10.1137/S0097539795293172}
\showDOI{\tempurl}
\showeprint{https://doi.org/10.1137/S0097539795293172}


\bibitem[\protect\citeauthoryear{Simon}{Simon}{1997}]%
        {Simon}
\bibfield{author}{\bibinfo{person}{Daniel~R. Simon}.}
  \bibinfo{year}{1997}\natexlab{}.
\newblock \showarticletitle{On the Power of Quantum Computation}.
\newblock \bibinfo{journal}{\emph{SIAM J. Comput.}} \bibinfo{volume}{26},
  \bibinfo{number}{5} (\bibinfo{year}{1997}), \bibinfo{pages}{1474--1483}.
\newblock
\urldef\tempurl%
\url{https://doi.org/10.1137/S0097539796298637}
\showDOI{\tempurl}
\showeprint{https://doi.org/10.1137/S0097539796298637}


\bibitem[\protect\citeauthoryear{Szegedy}{Szegedy}{2019}]%
        {szegedy2019qaoa}
\bibfield{author}{\bibinfo{person}{Mario Szegedy}.}
  \bibinfo{year}{2019}\natexlab{}.
\newblock \bibinfo{title}{What do {QAOA} energies reveal about graphs?}
\newblock
\newblock
\showeprint[arxiv]{1912.12277}~[quant-ph]


\bibitem[\protect\citeauthoryear{team and collaborators}{team and
  collaborators}{2020}]%
        {quantum_ai_team_and_collaborators_2020_4023103}
\bibfield{author}{\bibinfo{person}{Quantum~AI team} {and}
  \bibinfo{person}{collaborators}.} \bibinfo{year}{2020}\natexlab{}.
\newblock \bibinfo{booktitle}{\emph{qsim}}.
\newblock
\urldef\tempurl%
\url{https://doi.org/10.5281/zenodo.4023103}
\showDOI{\tempurl}


\bibitem[\protect\citeauthoryear{Terhal}{Terhal}{2018}]%
        {Terhal}
\bibfield{author}{\bibinfo{person}{Barbara~M. Terhal}.}
  \bibinfo{year}{2018}\natexlab{}.
\newblock \showarticletitle{Quantum supremacy, here we come}.
\newblock \bibinfo{journal}{\emph{Nature Physics}} \bibinfo{volume}{14},
  \bibinfo{number}{6} (\bibinfo{year}{2018}), \bibinfo{pages}{530--531}.
\newblock
\showISBNx{1745-2481}
\urldef\tempurl%
\url{https://doi.org/10.1038/s41567-018-0131-y}
\showDOI{\tempurl}


\bibitem[\protect\citeauthoryear{Tucci}{Tucci}{1995}]%
        {tucci1995quantum}
\bibfield{author}{\bibinfo{person}{Robert~R Tucci}.}
  \bibinfo{year}{1995}\natexlab{}.
\newblock \showarticletitle{Quantum {Bayesian} nets}.
\newblock \bibinfo{journal}{\emph{Int. Journal of Modern Physics B}}
  \bibinfo{volume}{9}, \bibinfo{number}{03} (\bibinfo{year}{1995}),
  \bibinfo{pages}{295--337}.
\newblock


\bibitem[\protect\citeauthoryear{van Dam, Hallgren, and Ip}{van Dam
  et~al\mbox{.}}{2003}]%
        {10.5555/644108.644189}
\bibfield{author}{\bibinfo{person}{Wim van Dam}, \bibinfo{person}{Sean
  Hallgren}, {and} \bibinfo{person}{Lawrence Ip}.}
  \bibinfo{year}{2003}\natexlab{}.
\newblock \showarticletitle{Quantum Algorithms for Some Hidden Shift Problems}.
  In \bibinfo{booktitle}{\emph{Proceedings of the Fourteenth Annual ACM-SIAM
  Symposium on Discrete Algorithms}} (Baltimore, Maryland)
  \emph{(\bibinfo{series}{SODA '03})}. \bibinfo{publisher}{Society for
  Industrial and Applied Mathematics}, \bibinfo{address}{USA},
  \bibinfo{pages}{489–498}.
\newblock
\showISBNx{0898715385}


\bibitem[\protect\citeauthoryear{Van Den~Nes}{Van Den~Nes}{2010}]%
        {VanDenNes:2010:CSQ:2011350.2011356}
\bibfield{author}{\bibinfo{person}{Maarten Van Den~Nes}.}
  \bibinfo{year}{2010}\natexlab{}.
\newblock \showarticletitle{Classical Simulation of Quantum Computation, the
  {Gottesman-Knill} Theorem, and Slightly Beyond}.
\newblock \bibinfo{journal}{\emph{Quantum Info. Comput.}} \bibinfo{volume}{10},
  \bibinfo{number}{3} (\bibinfo{date}{March} \bibinfo{year}{2010}),
  \bibinfo{pages}{258--271}.
\newblock
\showISSN{1533-7146}
\urldef\tempurl%
\url{http://dl.acm.org/citation.cfm?id=2011350.2011356}
\showURL{%
\tempurl}


\bibitem[\protect\citeauthoryear{{Yeang}}{{Yeang}}{2010}]%
        {5708827}
\bibfield{author}{\bibinfo{person}{C. {Yeang}}.}
  \bibinfo{year}{2010}\natexlab{}.
\newblock \showarticletitle{A Probabilistic Graphical Model of Quantum
  Systems}. In \bibinfo{booktitle}{\emph{2010 Ninth International Conference on
  Machine Learning and Applications}}. \bibinfo{pages}{155--162}.
\newblock
\urldef\tempurl%
\url{https://doi.org/10.1109/ICMLA.2010.30}
\showDOI{\tempurl}


\end{thebibliography}

\end{document}